\documentclass[aps,pra,twocolumn,footinbib, superscriptaddress, nofootinbib]{revtex4-1}

\usepackage{amssymb, amsmath}
\usepackage{graphicx} 
\usepackage{hyperref} 
\usepackage[utf8]{inputenc}
\usepackage[caption=false, position=top]{subfig}
\usepackage{textcomp} 
\usepackage{enumitem} 

\usepackage{orcidlink}
\begin{document}

\title{Impact of Annealing and Nanostructuring on Properties of \\ NV Centers Created by Different Techniques}

\author{Miriam Mendoza Delgado}
\thanks{These authors contributed equally.}
\affiliation{Institute of Nanostructure Technologies and Analytics (INA), University of Kassel, Heinrich-Plett-Strasse 40, 34132 Kassel, Germany}

\author{Lucas Tsunaki  \orcidlink{0009-0003-3534-6300}}
\thanks{These authors contributed equally.}
\affiliation{Department Spins in Energy Conversion and Quantum Information Science (ASPIN), Helmholtz-Zentrum Berlin für Materialien und Energie GmbH, Hahn-Meitner-Platz 1, 14109 Berlin, Germany}

\author{Shaul Michaelson}
\affiliation{Schulich Faculty of Chemistry, Technion - Israel Institute of Technology, Haifa 3200003, Israel}

\author{Mohan K. Kuntumalla}
\affiliation{Schulich Faculty of Chemistry, Technion - Israel Institute of Technology, Haifa 3200003, Israel}

\author{Johann P. Reithmaier}
\affiliation{Institute of Nanostructure Technologies and Analytics (INA), University of Kassel, Heinrich-Plett-Strasse 40, 34132 Kassel, Germany}

\author{Alon Hoffman \orcidlink{0000-0002-3100-9548}}
\affiliation{Schulich Faculty of Chemistry, Technion - Israel Institute of Technology, Haifa 3200003, Israel}

\author{Boris Naydenov \orcidlink{0000-0002-5215-3880}}
\email{boris.naydenov@helmholtz-berlin.de}
\affiliation{Department Spins in Energy Conversion and Quantum Information Science (ASPIN), Helmholtz-Zentrum Berlin für Materialien und Energie GmbH, Hahn-Meitner-Platz 1, 14109 Berlin, Germany}
\affiliation{Berlin Joint EPR Laboratory, Fachbereich Physik, Freie Universität Berlin, Arnimallee 14, 14195 Berlin, Germany}

\author{Cyril Popov \orcidlink{0000-0002-4646-5319}}
\email{popov@ina.uni-kassel.de}
\affiliation{Institute of Nanostructure Technologies and Analytics (INA), University of Kassel, Heinrich-Plett-Strasse 40, 34132 Kassel, Germany}

\date{\today}

\begin{abstract}
	
Nitrogen-vacancy (NV) centers in diamonds have been an epicenter of research for diverse applications in quantum technologies. It is therefore imperative that their fabrication techniques are well understood and characterized for the technological scalability of these applications. A comparative study of the optical and spin properties of NVs created by ion implantation and chemical vapor deposition delta-doping is thus presented, combined with an investigation on the impact of annealing in vacuum at different temperatures. In addition, nanopillars are fabricated by electron beam lithography and reactive ion etching for enhanced photon collection efficiency. An extensive combination of characterization techniques is employed. Notably, the smallest nanopillars present fluorescence enhancements of factor around 50, compared to the unstructured regions. Annealing is also demonstrated to increase the optical contrast between the NVs' electronic states, the coherence and relaxation times both in bulk as in pillars. Regarding the NV preparation technique, the delta-doping is shown to create NVs with less lattice defects and strain compared to implantation.

\end{abstract}

\maketitle

\section{Introduction}\label{sec:intro}

The NV center is a point defect in the diamond lattice composed of a nitrogen atom replacing a carbon atom adjacent to a vacancy, featuring unique optical and spin properties. A groundbreaking demonstration in the early 2000s showed that the electron spin of NV centers could be optically manipulated and read out at room temperature \cite{F.Jelezko2}, marking the beginning of their implementation in quantum sensing \cite{sensing_NV1, sensing_NV2, sensing_NV3}, communications \cite{communication_NV1} and computing \cite{F.Jelezko, computing_NV1, computing_NV2}. Subsequent research demonstrated the ability of NV centers to measure weak magnetic fields at the nanoscale, highlighting their potential in quantum magnetometry and nanoscale sensing \cite{Balasubramanian, sensor}. All these findings motivate further research on the NV properties and the precise characterization of its preparation techniques, in order to reach technological scalability for these applications.

Depending on the application, different techniques are preferred for creation of NV centers. For instance, nitrogen implantation is favored for creating single NV centers used in quantum computing, since it offers precise control over their depth and positioning \cite{ArifulHaque, FedericoGorrini}. Chemical vapor deposition (CVD) is commonly preferred for creating NV-rich diamonds for sensing and imaging applications \cite{TLuo}. Furthermore, heavily doped nitrogen delta layers are created by microwave plasma-assisted CVD with a rapid gas switching and a laminar gas flow to create sharp interfaces between doped and undoped diamond layers with thicknesses down to 2-3~nm \cite{deltaNVVikharev, deltaNVBogdanov}. Each of these methods has advantages and disadvantages. In the case of ion implantation, a precise control over the concentration and localization of the implanted ions can be achieved varying the energy and dose of implanted ions, which affect the NV properties. The implantation process, however, causes significant damage of the diamond lattice, deteriorating the relaxation times of the NV centers \cite{Jelezko3}.  One possible solution to minimize the lattice damage would be to keep the sample at high temperature during implantation. Thus, the defects created by the ions could diffuse and annihilate when reaching the diamond surface. Contrastingly in CVD NV-diamonds, the NV centers have better coherence properties, in the expense of a lower control over their localization and concentration \cite{AcostaHemmer}. In addition to these two mostly used techniques, NVs can be also created by direct laser writing as recently shown \cite{laser_writting}.

Another important point for applications in quantum technology is the incorporation of NV centers in photonic structures for enhancement of improving photon collection efficiency \cite{nanowire}, which is one of the requirements for a high-fidelity NV state readout \cite{DominikIrber}. Finally, annealing also plays a crucial role for the NV centers creation and their properties \cite{annealingLPHT, annealing1, annealing2} not only by improving the conversion efficiency of P1 centers to NVs, but also changing the local charge distribution and the diamond lattice strain around the NVs by enabling diffusion of charged defects \cite{charge_diffusion}  and relieving the lattice strain.

In order to characterize all these effects, we present the results of a comparative study of the optical and spin properties of NV centers created in diamond by ion implantation and CVD $\delta$-doping. In addition, the impacts of the nanostructuring of pillars with nominal diameters of 100 to 300~nm and of the annealing at 1200$^\circ$C and 1400$^\circ$C on these properties were investigated by a pool of techniques including photoluminescence spectroscopy, confocal fluorescence mapping and saturation curves, optically detected magnetic resonance (ODMR) of the NVs electronic spin, and spin echo measurements of the coherence and relaxation times of the NVs. 
 
\section{Materials and Methods}\label{sec:materials}

\subsection{CVD $\delta$-doping}\label{sec:Overgrowth}

Nitridation process with pure N$_{2}$ plasma in a microwave plasma (MWP) CVD setup was applied, followed by a short in situ diamond CVD overgrowth to create a nitrogen $\delta$-doped layer \cite{technion, technion2}. Electronic grade diamond 4.0 $\times$ 4.0 $\times$ 0.5~mm (Element Six) was used as a substrate. It was initially exposed to MWP-H cleaning of 30~min at 6~kW plasma power and 100~Torr (133.32~mbar) working pressure. Then the plasma was changed to pure N$_{2}$ (99.999\%) and treated for 30 min at 10 Torr (13.33~mbar)  and 1 kW. After that, the chamber was evacuated and diamond overgrowth was performed for 10~min at 100~Torr (133.32 mbar), 6~kW, 2\% CH$_{4}$ in H$_{2}$, and 950$^\circ$C substrate temperature. The thickness of the overgrown layer was about 800~nm considering the typical growth rate at the above mentioned conditions. Previous studies have shown the formation of N delta layer in diamond as revealed by sharp SIMS profiles of CN$^{-}$ in the spectra \cite{technion, technion2}.

\subsection{Ion Implantation}

Two implanted electronic grade membranes with thickness of 40~$\mu m$, one of them with (100) orientation, the other with (111), both with root mean square surface roughness of less than 1~nm (Qnami) were investigated for comparison. The ion implantation was performed with ion energy of 200~keV, resulting in an estimated NV depth of 200~nm and an expected density of over 1000~NVs/$\mu$m$^{2}$.


\subsection{Fabrication of Diamond Nanopillars} \label{sec:pillar_fabrication}
\begin{figure}[t!]
    \centering
    \includegraphics[width=\columnwidth]{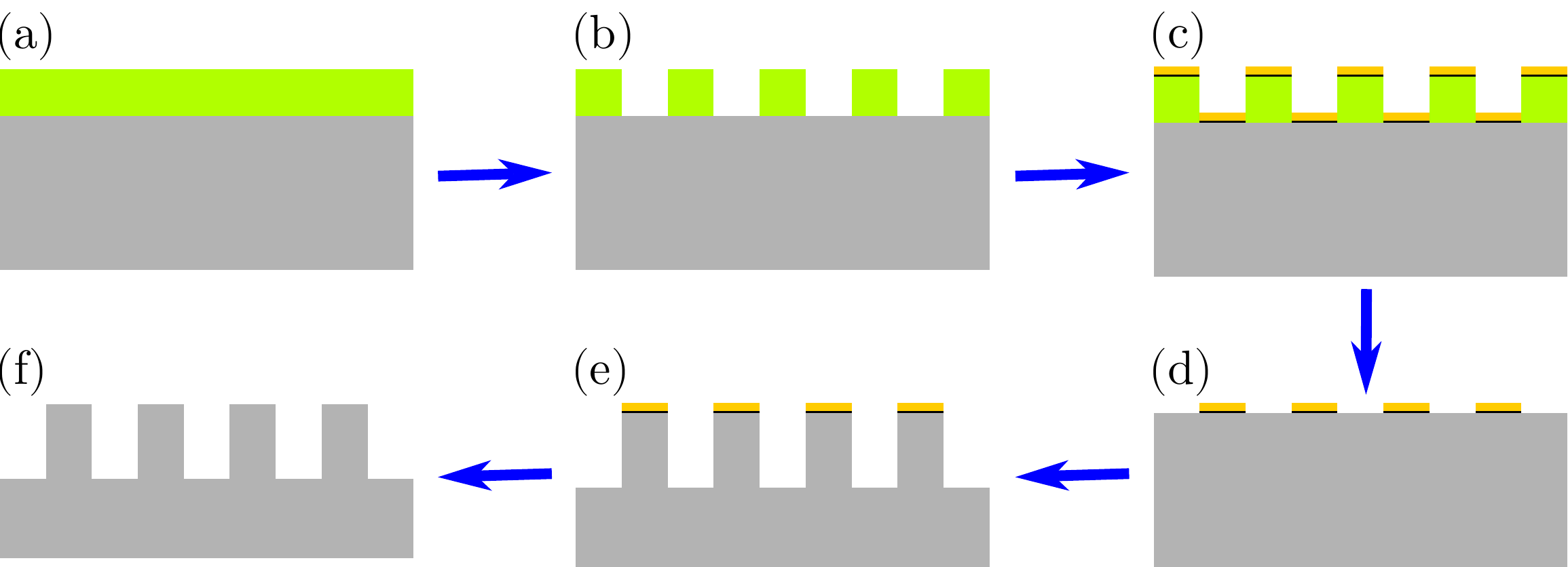} 
    \caption{Fabrication steps of diamond nanopillars: (a) Spin coating of resist (green) on diamond (gray); (b) Patterning of resist via eBeam lithography; (c) Deposition of Ti (black) and Au (yellow); (d) Lift-off in DMSO; (e) ICP-RIE with O2 plasma of diamond; (f) Mask removal.}
    \label{fig:fabrication_np}
\end{figure}

\begin{figure*}[t!]
	\centering
	\subfloat[(100)-implanted]{
		\includegraphics[height=4cm]{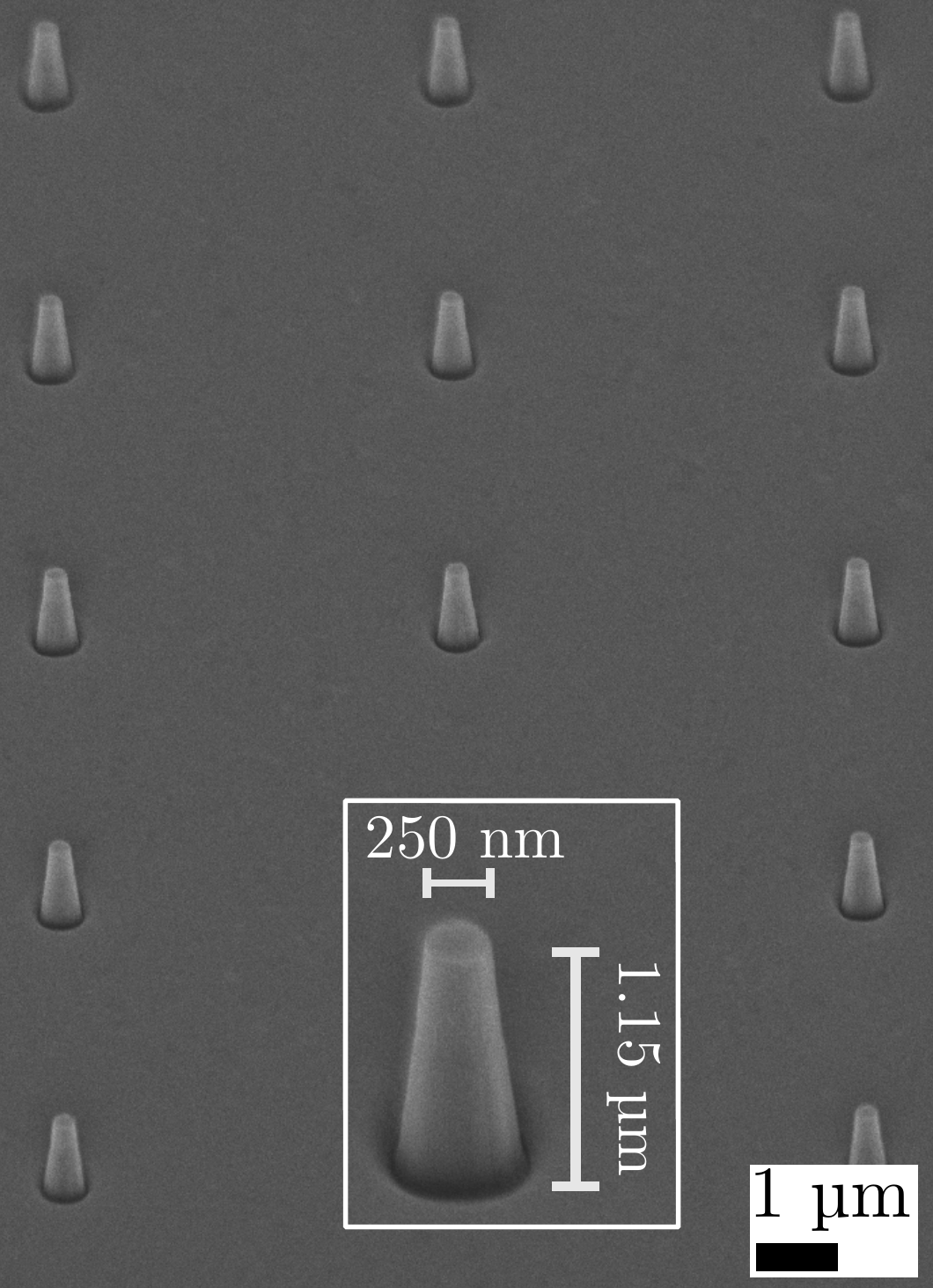}
		\label{SEM Qnami 100}
	}
	\hspace{0.1cm} 
	\subfloat[(111)-implanted]{
		\includegraphics[height=4cm]{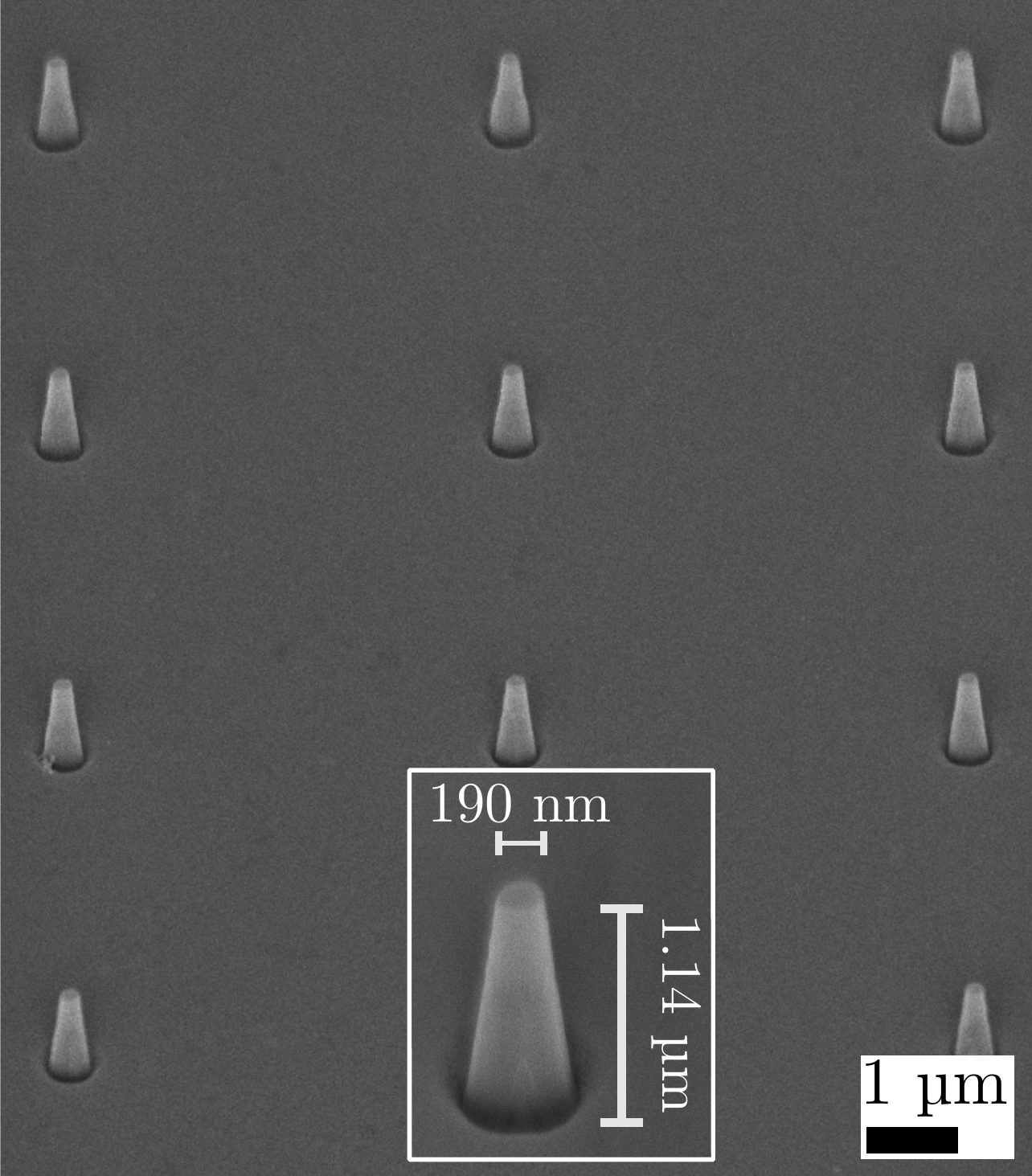}
		\label{SEM Qnami 111}
	}
	\hspace{0.1cm} 
	\subfloat[(111)-implanted]{
		\includegraphics[height=4cm]{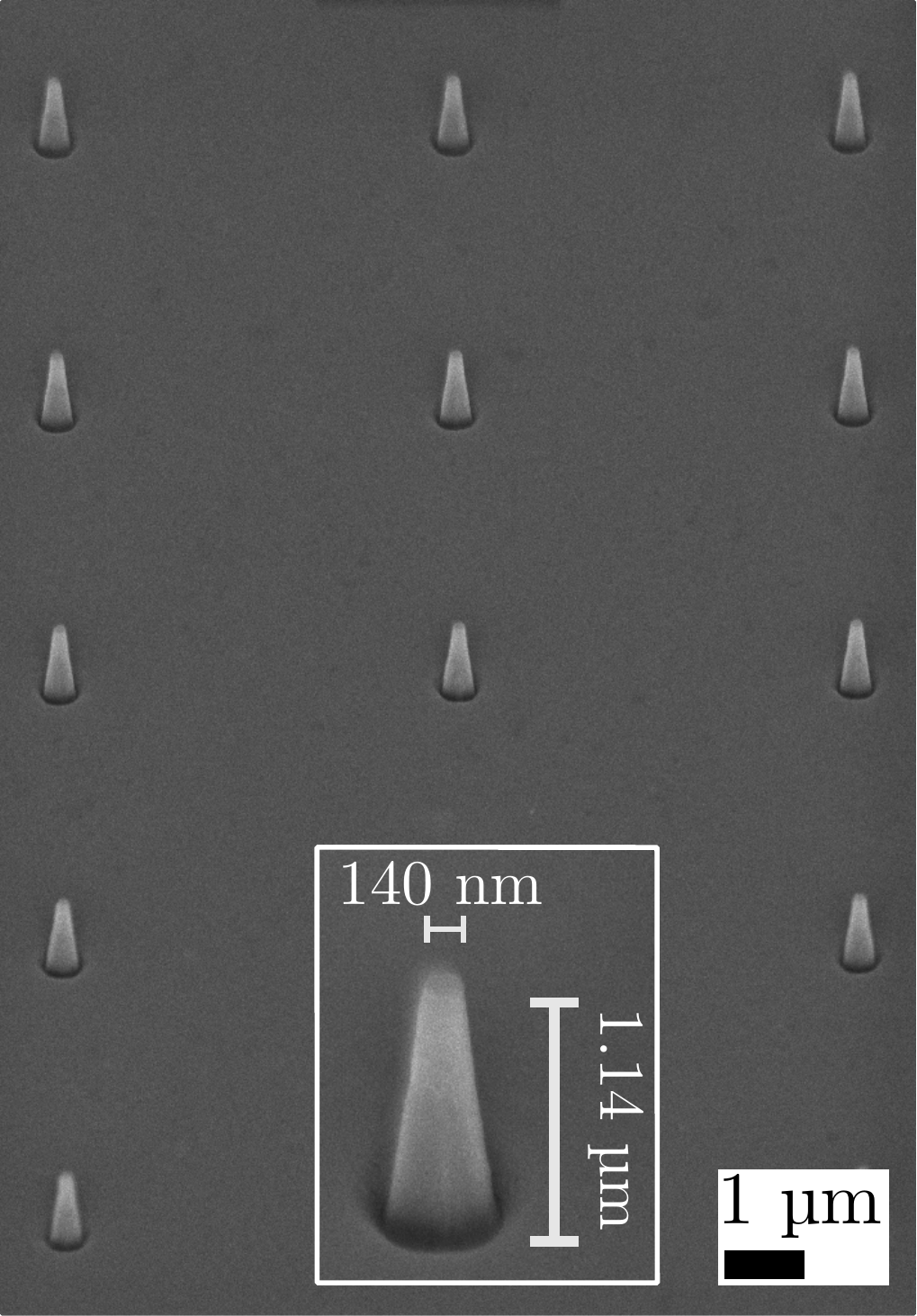}
		\label{SEM Qnami 111 DE}
	}
	\hspace{0.1cm} 
	\subfloat[$\delta-$doped]{
		\includegraphics[height=4cm]{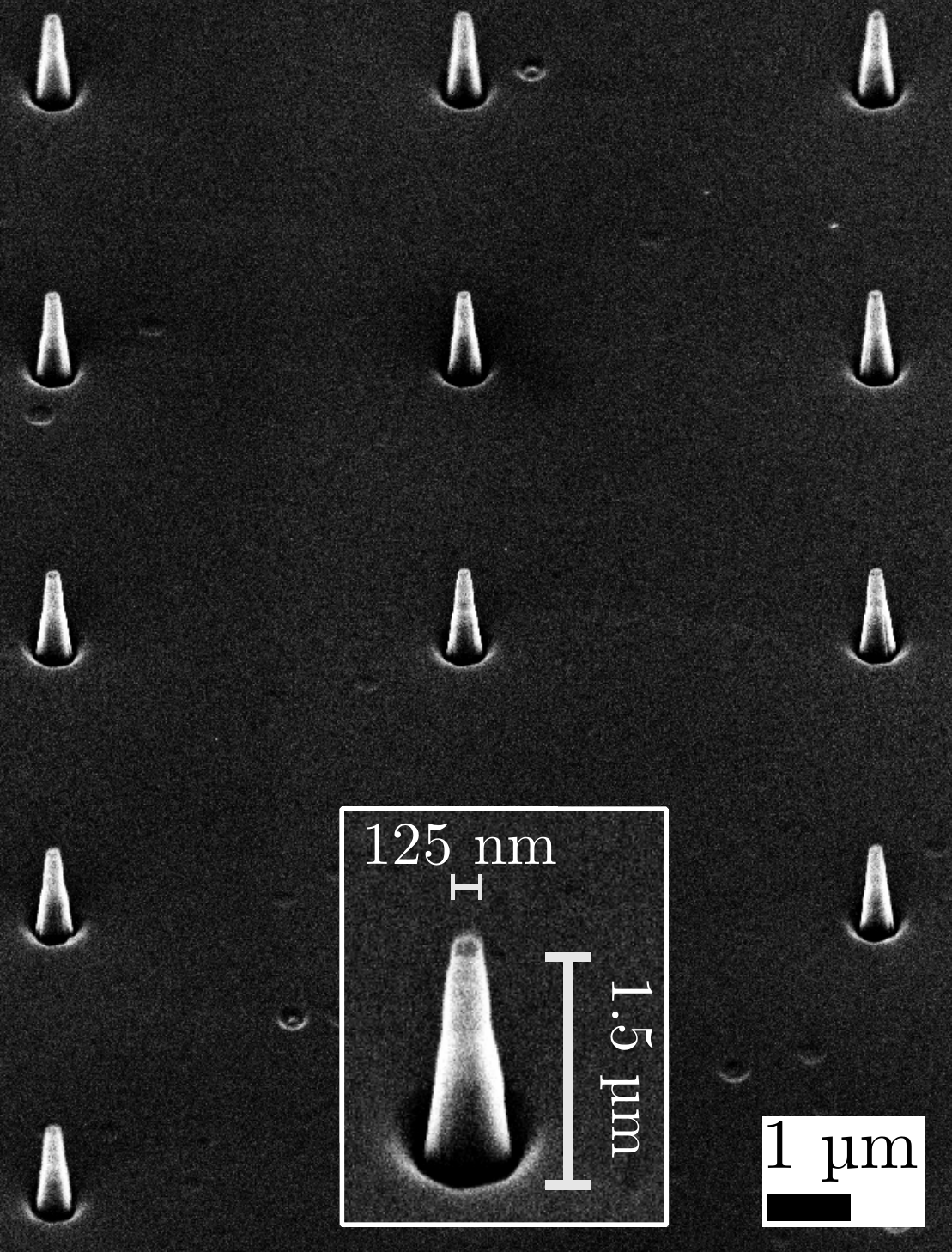}
		\label{SEM delta}
	}
	\caption{Typical SEM images of diamond nanopillar arrays taken at a tilt angle after removal of the metal hard masks: (a) (100)-implanted, nominal diameter 250~nm; (b) (111)-implanted, nominal diameter 200~nm; (c) (111)-implanted nominal diameter 150~nm (d) $\delta$-doped, nominal diameter 100~nm. The insets provide close-ups of individual pillars, displaying their calculated heights and measured top diameters.}
	\label{SEM nanopillars}
\end{figure*}

Nanopillars were fabricated on half of each of the three samples to enable a comparative study of the properties of incorporated NV centers in both bulk and structured areas. Electron beam (eBeam) lithography was utilized to create metal masks required for diamond structuring using inductively-coupled plasma - reactive ion etching (ICP-RIE). The nanopillars featured diameters ranging from 100~nm to 300~nm, which determined the size of the NV ensembles, i.e. the number of NVs incorporated in each nanopillar, and a center-to-center distance of 5~$\mu m$. The complete fabrication process is schematically represented in Fig. \ref{fig:fabrication_np}, additional details are provided in the Supplementary Material (Sec. \ref{sec:SMfabDNP}).

As a first step, eBeam lithography was employed to transfer the pattern with the arrays into PMMA resist [Fig. \ref{fig:fabrication_np} (a)]. To prevent charging effects during eBeam lithography, a conductive layer was spin-coated on top. After eBeam irradiation [Fig. \ref{fig:fabrication_np} (b)], the samples were immersed subsequently in developer and in stopper. To remove any residual resist in the developed areas, a descumming process was performed. In the next step, metal deposition was carried out with 10~nm of Ti as adhesion layer, followed by 200~nm Au [Fig. \ref{fig:fabrication_np} (c)]. In the subsequent step, the samples were immersed in DMSO to dissolve the underlying resist layers, leaving metal only in the areas without resist [Fig. \ref{fig:fabrication_np} (d)]. Finally, nanopillars were fabricated in the diamond samples by ICP-RIE with O$_{2}$ plasma [Fig. \ref{fig:fabrication_np} (e)]. After the etching, the gold mask was first chemically stripped, followed by the removal of the titanium adhesion layer [Fig. \ref{fig:fabrication_np} (f)]. The etch rate was estimated to be between 100 and 140~nm/min, monitored by measuring the height of the structures using scanning electron microscopy (SEM). By variation of the etching duration nanopillars with heights up to 1.6~$\mu m$ were fabricated.

\subsection{Annealing Under Vacuum Condition}

Previous studies have shown plasma-induced lattice damages and stress after ICP-RIE with O$_{2}$ plasma for diamond structuring, resulting in reduction of the photoluminescence emission and degradation of the spin properties of NV centers located in the vicinity of the etched surface \cite{ICPdamageNVnearSurface,ICPdamageNVnearSurface2, ICPdamageNVnearSurface3}. Annealing of the samples can improve the properties of NV centers, as demonstrated by Meng et al. \cite{annealingLPHT}. The diamond samples were individually annealed at pressures below $5\times10^{-6}$~mbar after nanostructuring in a home-made furnace. Details of the annealing protocols can be found in Sec. \ref{sec:SMannealing}.


\subsection{Characterization of Diamond Samples}
The properties of the NV centers were characterized in a custom-made confocal microscope, as similarly described in \cite{setup_1, setup_2}. The NV ensembles were optically excited with a scanning green laser of 518~nm wavelength focused on the diamond through an air objective, also used to collect the red fluorescence to determine the $m_s$ state. This resulted in a laser spot diameter of roughly 660~nm \cite{nanopillars_kseniia}. The fluorescence was then filtered, detected by an avalanche photo diode and counted. Alternatively, in photoluminescence experiments, the fluorescence was detected by a spectrometer. In saturation curve measurements, the laser power was varied and measured with a powermeter sensor, while in the other experiments the laser power was kept at 0.5~mW for efficient polarization of the $m_s$ state. 

Microwave fields were generated either by a continuous source or by an arbitrary waveform generator, depending on the experiment. In all cases, the microwave was amplified and transmitted to the NVs through a thin copper wire of 20~$\mu$m diameter. For the laser spot size under consideration, the resulting microwave field $\Vec{B}_1(t)$ was homogeneous in good approximation. A static magnetic field $\Vec{B}_0$ was applied by a permanent magnet resulting in a small intensity of about $|\Vec{B}_0| \approx  3.6$ mT, which was also homogeneous within the laser spot size. Finally, Qudi software \cite{qudi} was used for hardware control and data acquisition.


\section{Results and Discussion}\label{sec:results}


\subsection{Nanostructured Diamond Pillars}\label{sec:SEM}
The shape and dimensions of the nanopillars were primarily characterized by SEM. Fig. \ref{SEM nanopillars} displays typical SEM images of diamond nanopillar arrays from each sample taken at a tilt angle. In these images, the hard mask has been removed following the etching process, revealing the flat tops of the pillars. Each SEM image includes an inset that provides a close-up view of a single nanopillar, allowing for a detailed examination of its features. In the inset, the calculated height of the pillar is displayed together with the top diameter. The resulting average diameters of the pillars, obtained by measuring the top diameter of 15 different nanopillars in the same region, along with their standard deviations, are shown in Table \ref{tab:pillar_diameter} in Supplementary Material.

The fabrication process for the diamond nanopillar arrays, as discussed in Sec. \ref{sec:pillar_fabrication}, demonstrates high reproducibility across the samples, irrespective of their surface roughness and planarity. The pillars exhibited well-defined shapes, tapered with top diameters close to the nominal ones (between 100 and 300~nm) with standard deviation of less than 20~nm (except for some of the smallest pillars) and larger bottom diameters. 

The heights of the nanopillars structured in the implanted samples were very similar due to the same etching duration, namely 1.15~$\mu$m for the (100)-implanted sample and 1.14~$\mu$m for the (111)-implanted one. The prolonged etching resulted in heights of the nanopillars of ca. 1.5~$\mu$m for the $\delta-$doped sample. Finally, it should be mentioned, that the area between the nanopillars featured a smooth even surface suggesting minimal micromasking during the fabrication process.


\subsection{Photoluminescence}\label{sec:PL}

\begin{figure}[t!]
    \centering
    \includegraphics[width=\columnwidth]{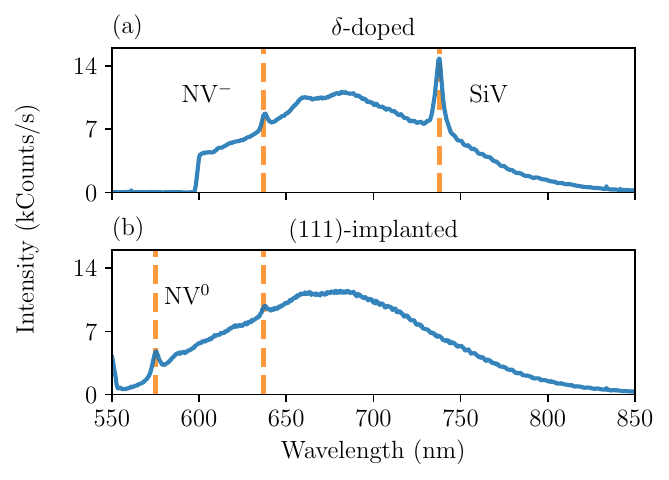}
    \caption{Photoluminescence measurements of (a) $\delta$-doped and (b) (111)-implanted samples, both presenting the ZPL emission from NV$^-$. The $\delta$-doped sample shows a strong SiV emission in addition to a strong Raman emission from diamond. For the implanted samples, it is possible to observe the ZPL from NV$^0$.}
    \label{fig:PL}
\end{figure}

Photoluminescence measurements were performed in the bulk region non structured area of all three diamond samples before annealing. Fig. \ref{fig:PL} (a) shows the spectrum from the $\delta$-doped sample and (b) from the (111)-implanted one. The (100)-implanted sample exhibits a spectrum equivalent to the (111) and thus is not shown here. Primarily, all samples have the characteristic zero-phonon line (ZPL) from NV$^-$ at 637.2~nm, indicating the effective creation of NV centers. In the implanted samples, the ZPL from NV$^0$ is also observed at 575~nm. On the other hand, the $\delta$-doped sample presents a strong Raman emission from the diamond host below 550~nm \cite{raman_diamond}, which saturates the counts at higher wavelengths, thus, requiring the use of 600~nm long pass filter blocking the NV$^0$ ZPL. Finally, an intense emission from silicon-vacancy (SiV) centers is also observed at 737~nm in the spectrum of the $\delta$-sample, which could be related to their creation from Si impurities during the CVD overgrowth of the sample.


\subsection{Fluorescence Mapping and Saturation Curves} \label{sec:sat}

\begin{figure*}[t!]
	\centering
	\includegraphics[width=.9\textwidth]{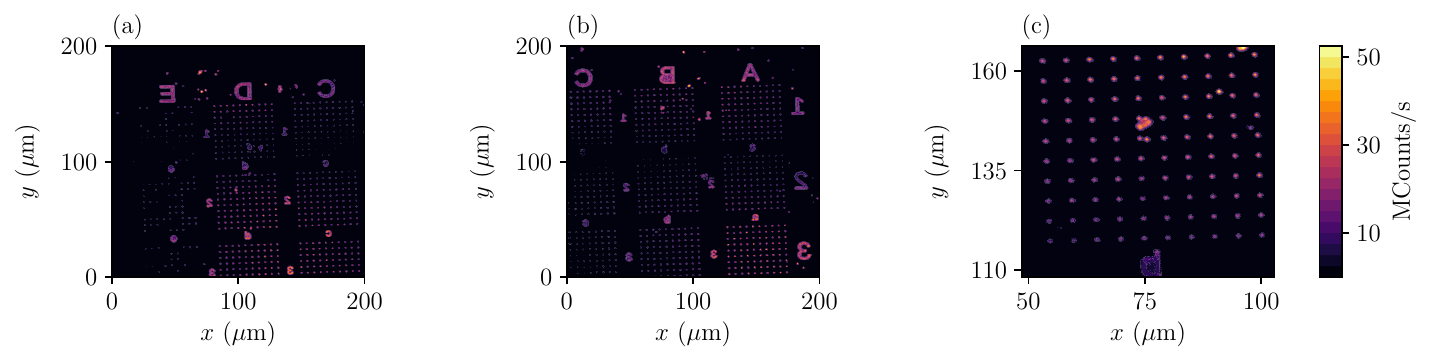}
	\caption{Fluorescence mapping of $\delta$-doped sample before annealing for (a) large scan areas showing regions E to C with small pillars, (b) regions C to A with larger pillars and (c) zoomed in the pillars from marker B area.}
	\label{fig:mapping}
\end{figure*}

The successful fabrication of the nanopillars with incorporated NV centers is further confirmed by the fluorescence mapping as shown in Fig.~\ref{fig:mapping} for the $\delta$-doped sample before annealing. Fig.~\ref{fig:mapping}~(a) and (b) show $200\times200$ $\mu$m$^2$ scans, where letter and number markers are visible, representing areas E to C and C to A, respectively, with different nominal pillar diameters given in Tab.~\ref{tab:pillar_diameter}. The shadow cast by the microwave wire antenna is also seen across $x$ at $y\approx100$ $\mu$m. Due to the large scanned area and the resulting tilt of the diamond surface, not all regions can be optically focused simultaneously and some regions appear brighter than others. On Fig.~\ref{fig:mapping}~(c), a smaller scale scan is presented over the B marker pillars, resulting in increased fluorescence. In addition, a small number of defects in the fabrication is also observed, as well as some missing pillars of the smallest diameter in region E. The other two samples show analogous fluorescence mapping before annealing.

\begin{figure*}[t!]
	\centering
	\includegraphics[width=\textwidth]{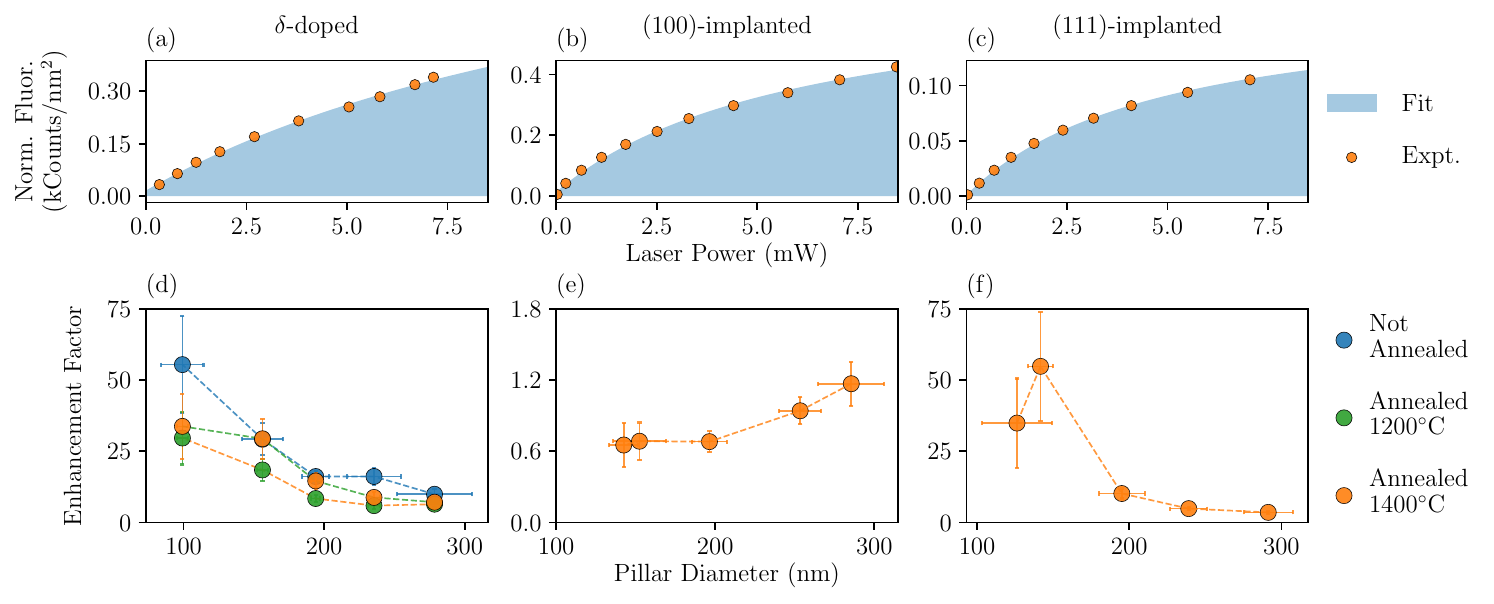}
	\caption{Area normalized saturation curves of the bulk region after annealing at 1400$^\circ$C in the (a) $\delta$-doped, (b) (100)-implanted and (c) (111)-implanted samples. The (111)-implanted sample has a fluorescence of about four times smaller than the other two, given by a less efficient NV fabrication yield. From the saturation curves, the enhancement factors of the pillars were calculated for the (d) $\delta$-doped, (e) (100)-implanted and (f) (111)-implanted samples. The $\delta$-doped and (111)-implanted samples show significant increase in the area normalized fluorescence of the pillars, which decreases with the pillar diameter, whereas the pillars in the (100)-implanted one have fluorescence comparable with the unstructured region.}
	\label{fig:enhan}
\end{figure*}

Even though the fluorescence mapping gives qualitative information about the photon collection efficiency properties of the photonic nanostructures, a more precise quantitative comparison could be obtained from fluorescence saturation curve measurements. In this case, the fluorescence counts are measured as a function of the laser power \cite{nanopillars_kseniia, sat_function} and then normalized by the area. The latter is determined by the laser spot size for the bulk (with an estimated diameter of 660~nm \cite{nanopillars_kseniia}) and by the diameters of the pillars measured by SEM. This way, saturation curves were obtained in the pillars and in the bulk regions, as shown in Fig.~\ref{fig:enhan}~(a), (b) and (c) for the bulk region after annealing at 1400$^\circ$C in the $\delta$-doped, (100)-implanted and (111)-implanted samples, respectively. The $\delta$-doped and (100)-implanted samples exhibit comparable fluorescence in the bulk, whereas the (111)-implanted one has fluorescence about four times smaller, which could be related to a less efficient NV yield by the implantation at this crystallographic orientation \cite{nanopillars_kseniia}.By the $\delta$-doped sample the saturation curves were measured before and after each annealing step. By the implanted samples a bleaching effect was observed before annealing, marked by a decay of the fluorescence over time, thus making these measurements unfeasible. Further saturation curves for the pillar regions of the three studied samples are shown in Figs.~\ref{fig:sat_delta} and \ref{fig:sat_implanted} in the Supplementary Material.

To calculate the enhancement factor from the pillars, the area normalized fluorescence was fitted as a function of the laser power according to the definition of P. Siyushev \textit{et al.} \cite{sat_function}
\begin{equation*}
    \mathcal{F}(P) = \mathcal{F}_\infty \frac{P}{P + P_{sat}} + C.
\end{equation*}
where $P$ is the laser power, $\mathcal{F}_\infty$ the fluorescence at saturation, $P_{sat}$ the saturation power and $C$ a constant offset from the background. The enhancement factor is given by  $\mathcal{F}_\infty$ parameter divided by the reference value of the unstructured bulk for the same annealing condition in each sample, then normalized by the pillar average area measured by SEM (Sec. \ref{sec:SEM}) or by the area of the laser spot (with an estimated diameter of 660~nm) in the case of the bulk region  \cite{nanopillars_kseniia}. Additionally, the uncertainty of the enhancement factor is calculated from error propagation of the fit uncertainties and experimental deviation of the pillars diameters.

The $\delta$-doped and (111)-implanted samples show significant increase (slightly higher by the first one) in the area normalized fluorescence of the pillars, reaching enhancement values around 50 for the smallest pillars, although they exhibit less intense absolute fluorescence than the larger ones as seen in Fig.~\ref{fig:mapping}.  As the pillars diameter increases the enhancement factors decrease, approaching the value of the unstructured bulk region. Conversely, in the (100)-implanted diamond the fluorescence in the smallest pillars is slightly below that of the bulk, and increases with the pillar diameter. This can be attribute to some graphitization of the sample caused by annealing, affecting the NV optical properties. The graphitization could be due to surface imperfections which vary for the different samples. Lastly, the annealing of the $\delta$-doped sample also induces a small decline in the enhancement factors of the smallest pillars. The inhomogeneity in the fluorescence enhancement factors between the different samples and diameters could be ascribed not only to surface defects leading to partial graphitization, but also to a great extent to the crystal quality. The overgrown nitrogen $\delta$-doped sample shows much stronger Raman signal (Sec.~\ref{sec:PL}) related to diamond, compared to the other two samples, suggesting a crystal lattice with less defects. After the annealing some of the defects in the implanted samples are removed, leading to an improved crystal quality of the pillars.


\subsection{Optically Detected Magnetic Resonance}\label{sec:ODMR}

\begin{figure*}[t!]
    \centering
    \includegraphics[width=\textwidth]{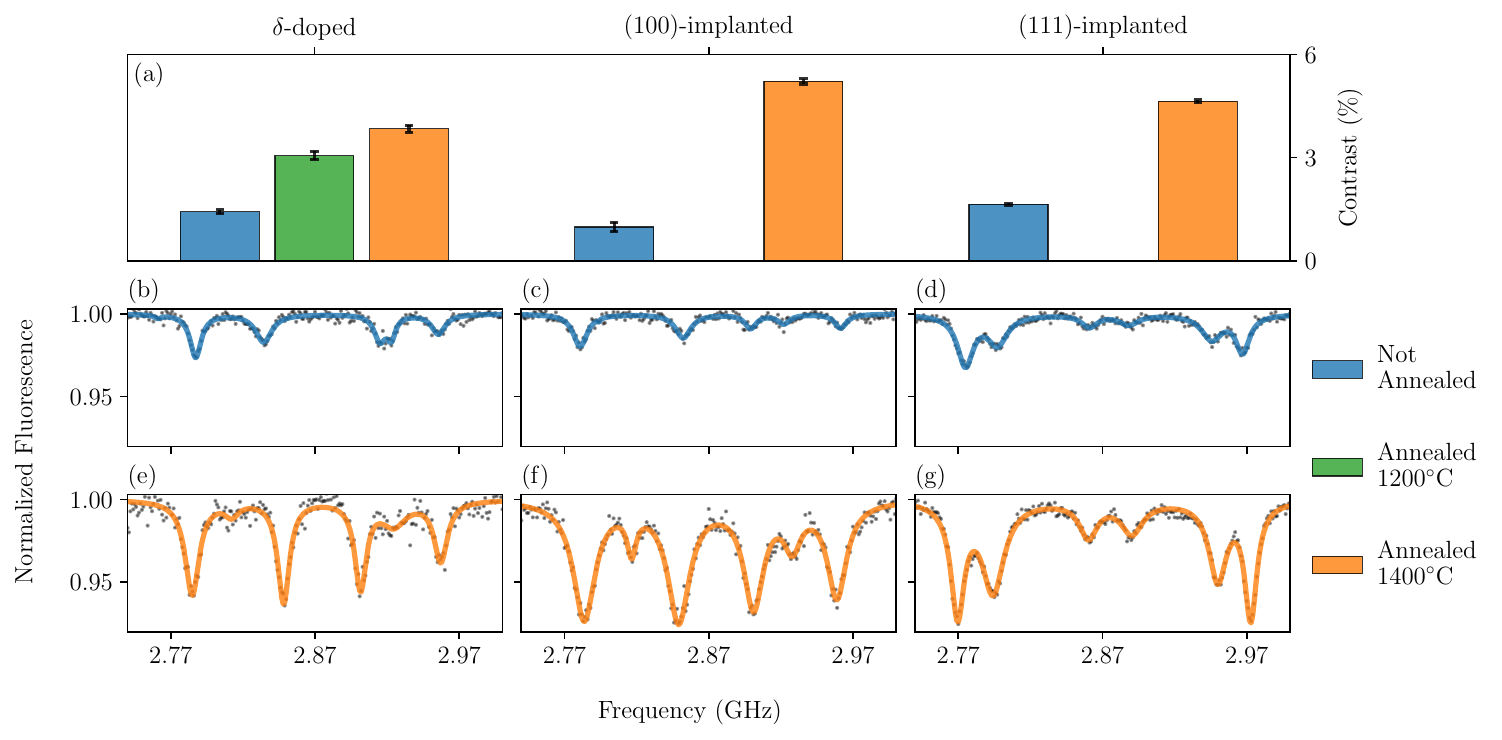}
    \caption{ODMR spectra before annealing of the bulk regions of the (b) $\delta$-doped, (c) (100)-implanted and (d) (111)-implanted samples; (e), (f) and (g) after annealing of the same samples. At least six resonances can be distinguished due to the different crystallographic orientations of the NVs, with different intensities in the (100) and (111) samples. (a) Annealing increases the average contrast of each sample, with the implanted samples having a larger contrast compared to the $\delta$-doped one.}
    \label{fig:ODMR}
\end{figure*}

Fig. \ref{fig:ODMR} presents the ODMR spectra measured before annealing in the bulk regions of the (b) $\delta$-doped, (c) (100)-implanted and (d) (111)-implanted samples, whereas (e), (f) (g) show the spectra from the same samples after annealing. All ODMR spectra were measured from similar distances of about 10~$\mu$m from the microwave antenna in order to have similar values of the MW magnetic field $\Vec{B}_1$. At least six resonance peaks can be distinguished in each spectrum, out of the eight expected resonances resulting from the four distinct crystallographic orientations of NVs, i.e. two overlap due to line-broadening. Furthermore, the different diamond orientations (100) and (111) exhibit dissimilar resonance contrast for each of the peaks at the same external magnetic field $\Vec{B}_0$.

In order to quantify the resonance contrast of each spectrum, the experimental data was fitted with a sum of six Lorentzian curves as 
\begin{equation*}
    \mathcal{L}_6(f) = C - \sum_{i=1}^{6} \frac{a_i \delta \omega_i^2}{(x-\omega_i)^2 + \delta \omega_i^2}.
\end{equation*}
with free parameter of $a_i$ as the intensity of each peak, $\delta \omega_i$ the width, $\omega_i$ the position and $C$ the total offset, typically 1. In turn, the total contrast of each spectrum as shown in Fig. \ref{fig:ODMR} (a) is defined by the average intensity $\sum_{i}a_i/6$, whereas the uncertainty is calculated from the covariance of the fit parameters. Notably, the contrast increases significantly with the annealing by a factor of almost three in the $\delta$-doped and (111)-implanted samples, and of 5.3 in the (100)-implanted one. The intermediate annealing step at 1200$^\circ$C of the $\delta$-doped sample also confirms this trend. In regards to the NV creation technique, the $\delta$-doped sample showed lower contrast enhancement after annealing than the implanted samples. This could be related to removal of some fluorescent defects and/or bettered stabilized NV charge state after the annealing \cite{annealing1}, better exhibited by implanted samples. 

The pillars show typically lower ODMR contrast than the bulk (data not shown), also with a significant increase after annealing. Nonetheless, the ODMR spectral contrast is not an appropriate quantitative metric for comparing the nanopillars. It is greatly influenced by the microwave field $\Vec{B}_1$, decaying with the distance from the microwave antenna, which in turn, is not equal for all pillars, given the geometric constrains of the experimental setup and the nanofabrication.


\subsection{Electron Spin Coherence}\label{sec:T2}

\begin{figure}[b!]
    \centering
    \includegraphics[width=\columnwidth]{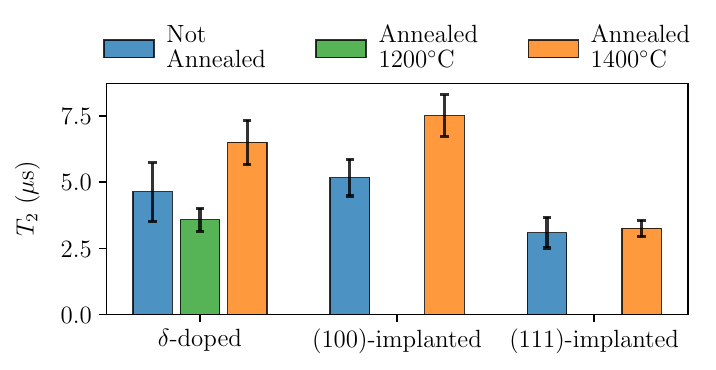}
    \caption{Coherence times of the bulk regions before and after annealing obtained from Hahn-echo experiments with the lowest frequency resonances. The (100)-implanted sample shows an increased coherence with annealing, while in the two other samples it remains the same within the uncertainty. The (100)-implanted and the $\delta$-doped samples have similar $T_2$, longer than that of the (111)-implanted one.}
    \label{fig:T2}
\end{figure}

\begin{figure*}[t!]
    \centering
    \includegraphics[width=\textwidth]{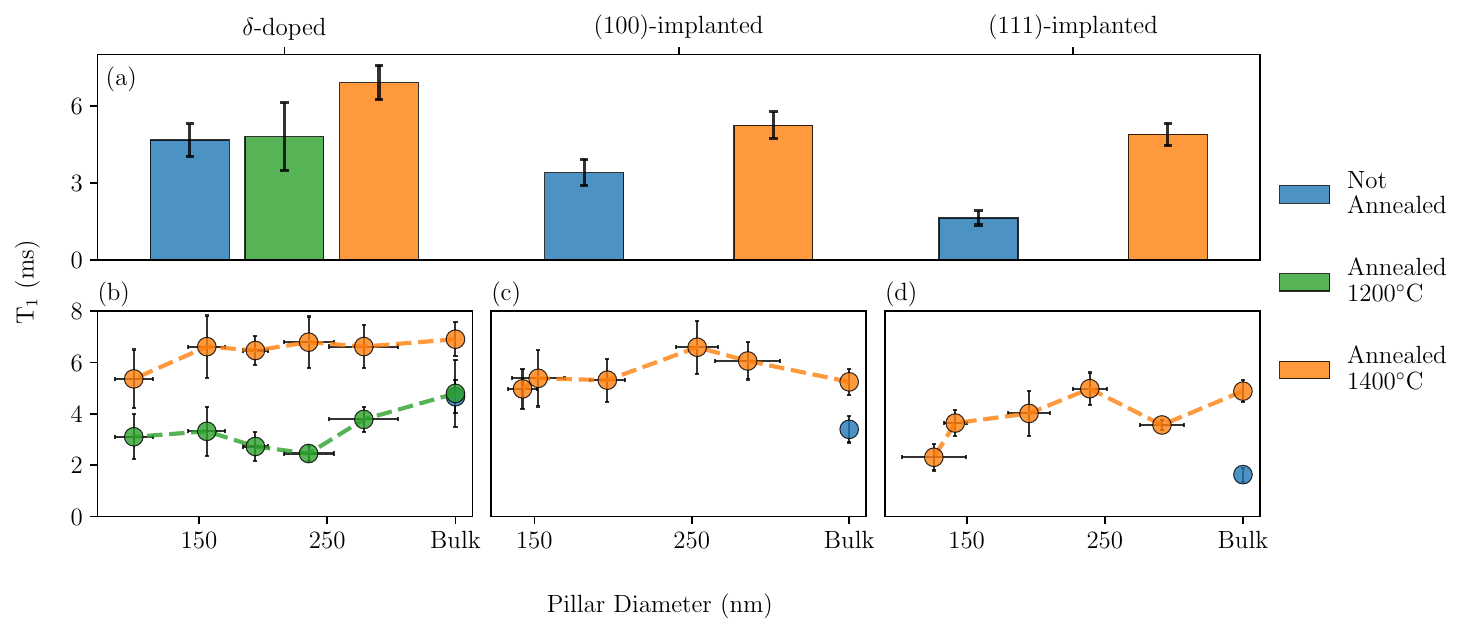}
    \caption{(a) Relaxation times $T_1$ of the bulk regions of the three samples before and after annealing. The $\delta$-doped sample shows slightly longer $T_1$ than the implanted ones, due to less lattice defects compared to the implantation. In addition, $T_1$ values show an increase after annealing, possibly resulting from a reduction of the lattice strain. (b) Annealing of the $\delta$-doped sample at higher temperature increases the relaxation times of the pillars, reaching the value of the bulk. In the implanted samples, $T_1$ of the bulk also increases with the annealing.}
    \label{fig:T1}
\end{figure*}

To measure the coherence times, a Hahn-echo sequence \cite{hahn} was applied to the electronic spin of the NVs, using the lowest ODMR resonant frequency from 2.77~GHz to 2.78~GHz. Rabi measurements \cite{rabi} were performed prior to the Hahn-echo to obtain the exact $\pi$-pulse duration. Finally, the $T_2$ times were obtained by fitting the Hahn-echo decay with an exponential function, with an error corresponding to the standard deviation from the fit parameter. The results for the bulk regions before and after annealing are shown in Fig.~\ref{fig:T2}.

The (111)-implanted sample has much shorter coherence than the (100)-implanted and $\delta$-doped samples which show similar $T_2$ values. A possible explanation could be the presence of some electron spin bath (for example P$_1$ centers with electron spin $S = 1/2$) in the diamond crystal even before the implantation which cannot be removed during the annealing step. The latter induced a small increase of $T_2$ in the $\delta$-doped sample (still within the uncertainty) and a definite increase by a factor of 1.4 in the (100)-implanted one. 

The coherence times in the pillars before annealing could not be measured due to insufficient optical contrast between the $m_s$ states in the $\delta$-doped sample and bleaching effects in the implanted ones. After annealing (data not shown), $T_2$ are comparable to the bulk with typically larger uncertainty, which could be related to magnetic noise coming from surface spins in the pillar walls.


\subsection{Electron Spin Relaxation}\label{sec:T1}

Relaxation times $T_1$ of the NV electron spin for the lowest frequency resonance were measured in the bulk regions of the three samples before and after annealing, as shown in Fig. \ref{fig:T1} (a). The $T_1$ values show an increase with annealing, being most noticeable in the (111)-implanted sample with an enhancement by a factor of 3. Furthermore, the $\delta$-doped sample has slightly longer relaxation time than the two other samples, but barely above uncertainty. This could be related to increased concentration of lattice defects caused by the ion implantation in the other two samples.

Another result comes from the comparison of the relaxation times of the pillars in the $\delta$-doped sample after the annealing steps at 1200$^\circ$C and 1400$^\circ$C, as shown in Fig. \ref{fig:T1} (b). With the lower temperature annealing, the smaller pillars show shorter $T_1$ than the bulk, increasing with the diameter due to less lattice strain. However, after the higher temperature annealing, the lattice strain is relaxed and they start to reach the value of the bulk region. A similar trend of an increase in $T_1$ with the pillar diameter is also observed in the (111)-implanted sample after the annealing at 1400$^\circ$C (Fig. \ref{fig:T1} (d)), whereas in the (100)-implanted one it remains essentially constant with the pillar size (Fig. \ref{fig:T1} (c)). Like by the Hahn echo experiments, the relaxation of the pillars before annealing could not be measured due to low contrast and bleaching effects.


\section{Conclusions} \label{sec:conclusion}

Color centers in diamond have a great potential in different fields, ranging from quantum sensing to quantum cryptography and communications. It is thus essential to optimize and precisely characterize different techniques for creation of NV, analyzing also the impact of annealing on the optical and spin properties of the NVs. We investigated comprehensively $\delta$-doped diamond via CVD overgrowth and two ion implanted samples with different crystal orientations. Comparing the bulk regions of the samples, $\delta$-doping showed longer relaxation times due to less lattice defects and strain, compared to the ion implantation. On the other hand, the $\delta$-doped sample exhibited a smaller ODMR contrast compared to the implanted samples. Regarding the crystal orientation, the two implanted samples have comparable relaxation and ODMR contrast, but the (111)-oriented sample shows much shorter coherence times. Photoluminescence spectroscopy evidenced the collateral creation of SiV centers due to Si impurities during the CVD overgrowth, whereas it was possible to observe the ZPL from NV$^0$ in addition to NV$^-$ in the implanted samples.

Apart from the NV creation techniques, it is of great interest for many applications to study the impact of diamond nanostructuring on the NV properties. In particular, pillars with diameters ranging from 100 to 300~nm were fabricated in each sample using electron beam lithography and reactive ion etching. The main interest on the fabrication of nanopillars lies in the improved photon collection efficiency (enhancement) compared to the flat surface bulk regions, which is substantiated by the measured saturation curves. Divergent behaviors were observed for each sample. In the (100)-implanted sample the fluorescence enhancement of the pillars was close to the unstructured region, in the other two samples strong enhancement with factors of around 50 were observed in the smallest pillars, decreasing with the pillar diameter. At the same time these pillars showed lower ODMR contrast and shorter $T_2$ times than the unstructured bulk region in the studied diamond samples. 

Other significant changes in the NV properties arise from the annealing of the samples. Whereas the coherence time slightly increased in the (100)-implanted sample and kept constant for the two other, the ODMR contrasts showed strong increase with the annealing in all samples, more pronounced in the (100)-implanted one by a factor of 5.3. The NV electron spin relaxation times in the unstructured regions also increased after annealing, most notably by a factor of 3.0 in the (111)-implanted sample. After an intermediate annealing step of the $\delta$-doped sample at 1200$^\circ$C, the pillars exhibited shorter $T_1$ times than the bulk, which increased with the pillar diameter. After the annealing at 1400$^\circ$C, the reduced lattice strain in the pillars resulted in saturation of the relaxation times, reaching the value of the bulk. Regarding the optical properties, the $\delta$-doped sample evidenced a small reduction of the fluorescence from the smaller pillars with the annealing, whereas the (100)-implanted one suffered from partial graphitization, impairing the collection efficiency from the nanopillars, but without significant effect on the spin properties of the NVs.

This work contributes to the ongoing advancement and characterization of the fabrication techniques of color centers in diamond and potentially other systems in quantum technologies. Specifically, these results constitute an initial step towards the preparation of recently proposed diamond based quantum token \cite{qtoken_1, qtoken_2}, aiming to use quantum states of NV ensembles in nanopillars as unclonable authentication keys. Further studies can concentrate on the very high fluorescence enhancements in the smallest pillars or the determination of the exact concentration of NV and P1 centers in differently prepared samples by double electron-electron resonance (DEER) techniques \cite{DEER}.

\section*{Acknowledgments}
    
We acknowledge Christos Thessalonikios and Kseniia Volkova from Helmholtz-Zentrum Berlin for their assistance on the experimental data acquisition for the characterization of the diamond samples. This work was supported by the German Federal Ministry of Education and Research (BMBF) under the projects DIQTOK (grant 16KISQ034) and DIAQUAM (grants 13N16956 and 13N16958). In addition, this work received funding from the German Research Foundation (DFG) under grants 410866378 and 41086656. S.M., M.K.K. and A.H. would like to thank the Technion-IIT for the promotion of research. This work was supported by the Israel Science Foundation (Grant No. 557/23). We would also like to thank Helen Diller Quantum Center, Technion for supporting financially this work. 

\setcounter{equation}{0}
\setcounter{figure}{0}
\setcounter{table}{0}
\setcounter{section}{0}

\renewcommand{\theequation}{S\arabic{equation}}
\renewcommand{\thefigure}{S\arabic{figure}}
\renewcommand{\thesection}{S\arabic{section}}
\renewcommand{\thetable}{S\arabic{table}}


\begin{table*}[t!]
    \centering
    \begin{tabular}{|c||c|c|c|c|c|}
    \hline
    Sample & A [nm] & B [nm] & C [nm] & D [nm] & E [nm] \\ \hline
    nominal diameters & 300 & 250 & 200 & 150 & 100 \\
    $\delta$-dopped & $280 \pm 30$ & $240 \pm 20$ & $190\pm 10$ & $160 \pm 10$ & $100 \pm 10$ \\
    (100)-implanted & $280 \pm 20$ & $250 \pm 10$ & $200\pm 10$ & $150 \pm 20$ & $142 \pm 9$ \\
    (111)-implanted & $290 \pm 20$ & $240 \pm 10$ & $190\pm 10$ & $142 \pm 8$ & $130 \pm 20$ \\
    \hline  
    \end{tabular}
    \caption{Average diameters of the nanopillars determined from the SEM images for each sample. Uncertainty is given by the standard deviation.}
    \label{tab:pillar_diameter}
\end{table*}

\section{Fabrication of Diamond Nanopillars} \label{sec:SMfabDNP}
\subsection{Fabrication of Metal Etching Masks}

As a first step, eBeam lithography was employed to transfer the pattern with the arrays of nanopillars into PMMA resist. To facilitate handling, the diamond samples were fixed on quartz glass pieces of $1\times1$~cm using wafer bond (HT-10.11, Brewer Science).Then they were cleaned using acetone and isopropanol, dried with nitrogen and dehydrated on a hot plate for 10~min at 120°C.
Two layers of PMMA (AR-P 617.06, Allresist GmbH) were consequently spin-coated for 60~s at 7000~rpm, resulting in a thickness of 225~nm each. The layers were baked at different temperatures: 200°C for the lower layer and 180°C for the upper one. 
This temperature difference enhances the sensitivity of the lower layer, allowing the developer to erode it faster and to create pronounced undercut structures that facilitate the lift-off process. To prevent charging effects during eBeam lithography, a conductive layer (Electra 92 AR-PC 5090, Allresist GmbH) was spin-coated on top and baked at 90°C for 60~s. After eBeam irradiation, the samples were rinsed with water and dried with nitrogen to remove the conductive layer. For development, the samples were immersed in developer (AR 600-50, Allresist GmbH) for 85~s, followed by immersion in stopper (AR 600-60, Allresist GmbH) for 30~s. To remove any residual resist, a descumming process was performed using oxygen plasma for 1~min at RF power of 100~W (TePla 200-G Oxygen Asher). 
In the next step, metal deposition was carried out on the samples using a high-vacuum evaporation system (Balzers BAK 600). Due to the poor adhesion of gold to diamond, an adhesion layer is necessary. To this end, 10~nm of Ti were deposited as adhesion layer, followed by a 200~nm layer of Au. Finally, the samples were immersed in DMSO at 80°C overnight to dissolve the underlying resist layers, leaving metal only in the areas without resist.

\subsection{Etching of Diamond Nanopillars}
Nanopillars, were fabricated in the diamond samples by ICP-RIE with  O$_{2}$
plasma (Oxford Plasmalab 100 ICP-RIE) applying the following parameters: 90~W
RF power, 1100~W ICP power, a pressure of 5~mTorr, a flow rate of 50~sccm O$_{2}$ , and a
substrate temperature of 20°C. After the etching, the gold mask was first stripped
using a potassium iodide solution (KI/I$_{2}$ 4:1) for 5~min, followed by the removal of the
titanium adhesion layer with a solution of HF$^{*}$:H$_{2}$O$_{2}^{*}$:H$_{2}$O with a ratio of 1:1:20, where HF$^{*}$ denotes a 50\% HF solution in H$_{2}$O and H$_{2}$O$_{2}^{*}$ represents a 30\% H$_{2}$O$_{2}$ solution in H$_{2}$O.

\section{Annealing Under Vacuum Condition ($<~5\cdot10^{-6}$~mbar)}\label{sec:SMannealing}

\begin{figure}[h!]
	\centering
	\includegraphics{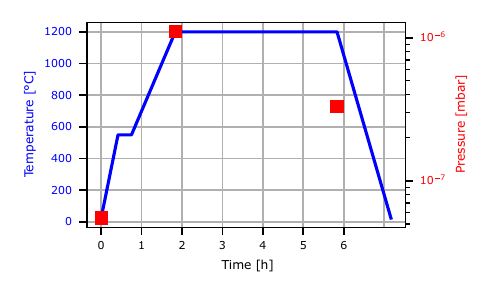}
	\caption{Temperature profile of the first annealing of the $\delta-$doped sample after
		nanopillar structuring. Pressure values measured at several annealing points are also
		given.}
	\label{SMannealing1}
\end{figure}

\begin{figure}[h!]
	\centering
	\includegraphics{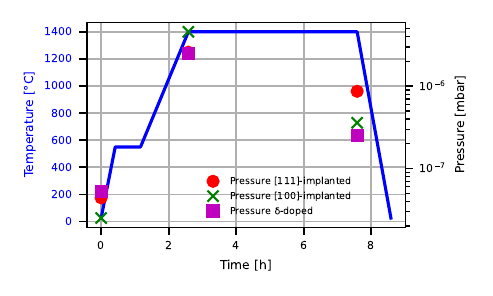}
	\caption{Temperature profile of the annealing of the (100) and (111)-implanted
		samples after nanopillar structuring and of the second annealing of the $\delta-$doped
		sample. Pressure values measured at several annealing points for each sample are
		also given.}
	\label{SMannealing2}
\end{figure}

The diamond samples under study were annealed after nanostructuring in a home-made furnace provided with a dual stage oil sealed rotary vane vacuum pump (TRIVAC NT5, Oerlikon Leybold Vacuum), a turbomolecular pump (TURBOVAC SL80, Leybold), a frequency converter (TURBO.DRIVE TD 400, Leybold), a two channels digital vacuum display and control unit (VD12, Thyracont), vacuum controller (Oerlikon Leybold Vacuum), an ultra-high vacuum ceramic heater stage (HTR-2-100, tectra) and a heater controller (HC3500, tectra). Prior to annealing, the samples were immersed in a wafer bond remover to detach them from the quartz glass holders and thoroughly cleaned with isopropanol, acetone, and piranha solution (H$_{2}$SO$_{4}$:H$_{2}$O$_{2}$ 3:1) to prevent any potential graphitization or contamination.

The $\delta-$doped sample was initially annealed at 1200°C after nanopillar fabrication with the temperature profile shown in Figure \ref{SMannealing1}. The first round of characterization, it was annealed at 1400°C with the temperature profile presented in Figure \ref{SMannealing2}. The same conditions were applied for the annealing of both implanted samples.

\begin{figure*}[h!]
	\centering
	\includegraphics[width=.9\textwidth]{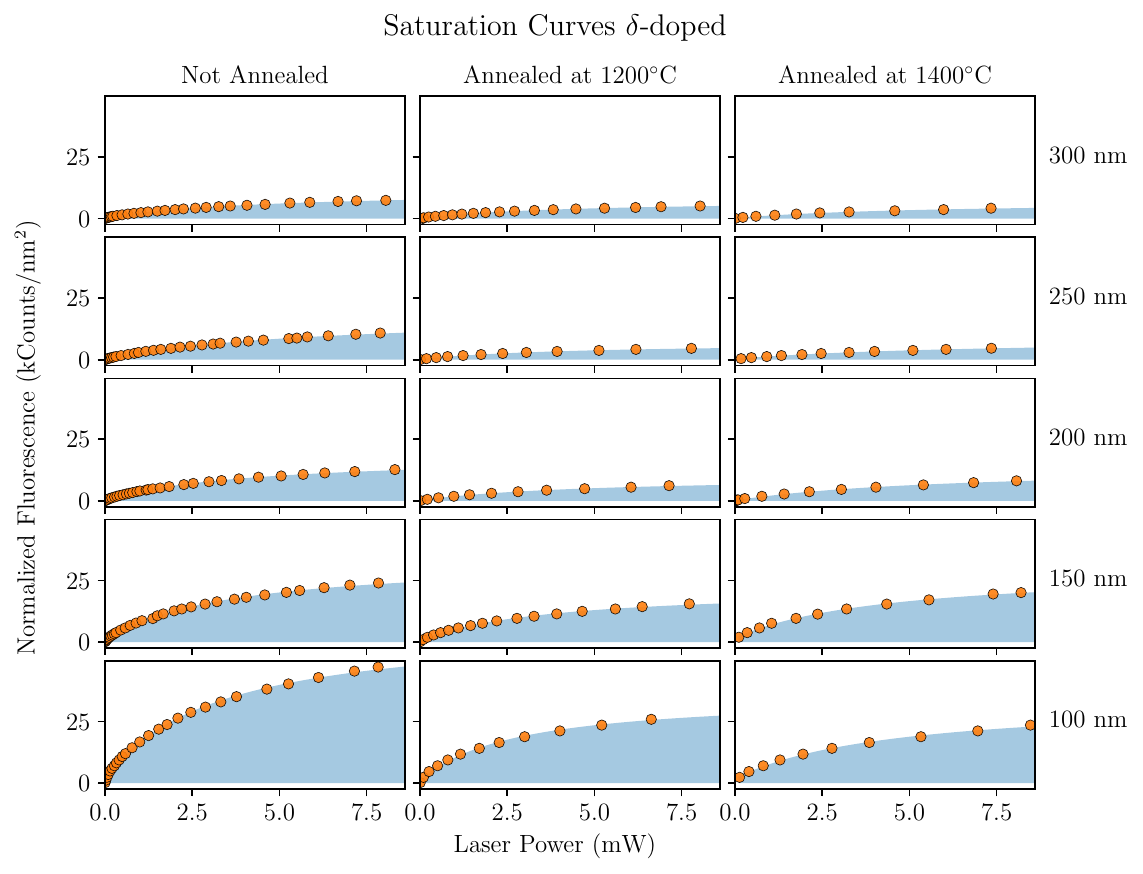}
	\caption{Area normalized saturation curves of $\delta$-doped sample measured in the pillars before and after each annealing step. The area normalized fluorescence shows a great increase in the smallest pillars, as also shown in Fig.~\ref{fig:enhan} (d).}
	\label{fig:sat_delta}
\end{figure*}

\begin{figure*}[h!]
	\centering
	\includegraphics[width=\columnwidth]{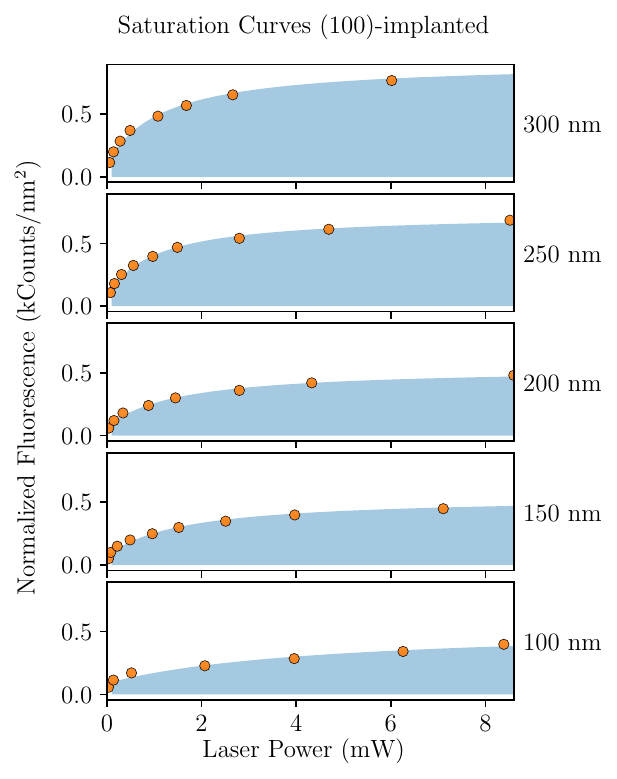}
	\hfill
	\includegraphics[width=\columnwidth]{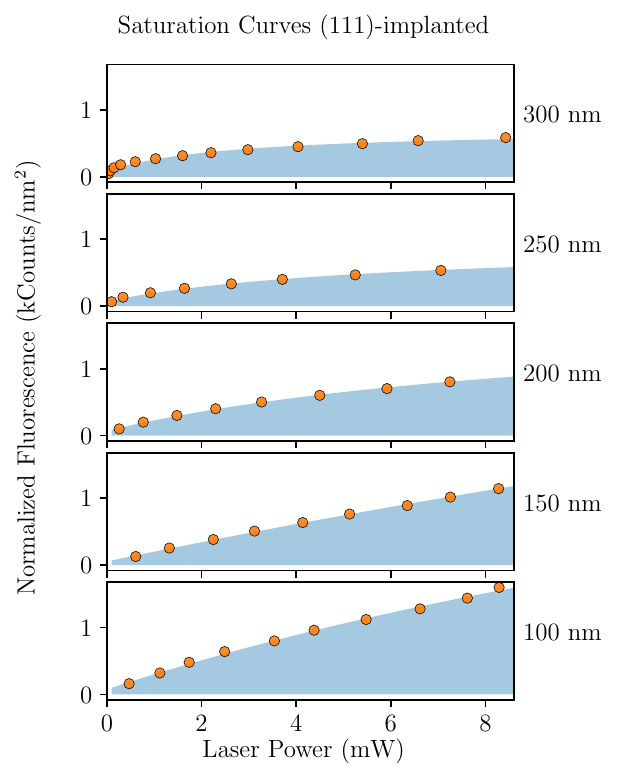}
	\caption{Area normalized saturation curves of the implanted sample measured in the pillars after annealing at 1400°C. The (111)-implanted has a small increase in fluorescence in the smaller pillars, while the (100)-shows a decrease.}
	\label{fig:sat_implanted}
\end{figure*}

\bibliography{references}

\begin{thebibliography}{40}%
\makeatletter
\providecommand \@ifxundefined [1]{%
 \@ifx{#1\undefined}
}%
\providecommand \@ifnum [1]{%
 \ifnum #1\expandafter \@firstoftwo
 \else \expandafter \@secondoftwo
 \fi
}%
\providecommand \@ifx [1]{%
 \ifx #1\expandafter \@firstoftwo
 \else \expandafter \@secondoftwo
 \fi
}%
\providecommand \natexlab [1]{#1}%
\providecommand \enquote  [1]{``#1''}%
\providecommand \bibnamefont  [1]{#1}%
\providecommand \bibfnamefont [1]{#1}%
\providecommand \citenamefont [1]{#1}%
\providecommand \href@noop [0]{\@secondoftwo}%
\providecommand \href [0]{\begingroup \@sanitize@url \@href}%
\providecommand \@href[1]{\@@startlink{#1}\@@href}%
\providecommand \@@href[1]{\endgroup#1\@@endlink}%
\providecommand \@sanitize@url [0]{\catcode `\\12\catcode `\$12\catcode
  `\&12\catcode `\#12\catcode `\^12\catcode `\_12\catcode `\%12\relax}%
\providecommand \@@startlink[1]{}%
\providecommand \@@endlink[0]{}%
\providecommand \url  [0]{\begingroup\@sanitize@url \@url }%
\providecommand \@url [1]{\endgroup\@href {#1}{\urlprefix }}%
\providecommand \urlprefix  [0]{URL }%
\providecommand \Eprint [0]{\href }%
\providecommand \doibase [0]{http://dx.doi.org/}%
\providecommand \selectlanguage [0]{\@gobble}%
\providecommand \bibinfo  [0]{\@secondoftwo}%
\providecommand \bibfield  [0]{\@secondoftwo}%
\providecommand \translation [1]{[#1]}%
\providecommand \BibitemOpen [0]{}%
\providecommand \bibitemStop [0]{}%
\providecommand \bibitemNoStop [0]{.\EOS\space}%
\providecommand \EOS [0]{\spacefactor3000\relax}%
\providecommand \BibitemShut  [1]{\csname bibitem#1\endcsname}%
\let\auto@bib@innerbib\@empty
\bibitem [{\citenamefont {Jelezko}\ \emph
  {et~al.}(2004{\natexlab{a}})\citenamefont {Jelezko}, \citenamefont {Gaebel},
  \citenamefont {Popa}, \citenamefont {Gruber},\ and\ \citenamefont
  {Wrachtrup}}]{F.Jelezko2}%
  \BibitemOpen
  \bibfield  {author} {\bibinfo {author} {\bibfnamefont {F.}~\bibnamefont
  {Jelezko}}, \bibinfo {author} {\bibfnamefont {T.}~\bibnamefont {Gaebel}},
  \bibinfo {author} {\bibfnamefont {I.}~\bibnamefont {Popa}}, \bibinfo {author}
  {\bibfnamefont {A.}~\bibnamefont {Gruber}}, \ and\ \bibinfo {author}
  {\bibfnamefont {J.}~\bibnamefont {Wrachtrup}},\ }\href {\doibase
  10.1103/PhysRevLett.92.076401} {\bibfield  {journal} {\bibinfo  {journal}
  {Physical Review Letters}\ }\textbf {\bibinfo {volume} {92}},\ \bibinfo
  {pages} {076401} (\bibinfo {year} {2004}{\natexlab{a}})}\BibitemShut
  {NoStop}%
\bibitem [{\citenamefont {Chen}\ \emph {et~al.}(2022)\citenamefont {Chen} \emph
  {et~al.}}]{sensing_NV1}%
  \BibitemOpen
  \bibfield  {author} {\bibinfo {author} {\bibfnamefont {S.}~\bibnamefont
  {Chen}} \emph {et~al.},\ }\href {\doibase 10.1073/pnas.2118876119} {\bibfield
   {journal} {\bibinfo  {journal} {Proceedings of the National Academy of
  Sciences}\ }\textbf {\bibinfo {volume} {119}},\ \bibinfo {pages}
  {e2118876119} (\bibinfo {year} {2022})}\BibitemShut {NoStop}%
\bibitem [{\citenamefont {Aslam}\ \emph {et~al.}(2023)\citenamefont {Aslam}
  \emph {et~al.}}]{sensing_NV2}%
  \BibitemOpen
  \bibfield  {author} {\bibinfo {author} {\bibfnamefont {N.}~\bibnamefont
  {Aslam}} \emph {et~al.},\ }\href {\doibase 10.1038/s42254-023-00558-3}
  {\bibfield  {journal} {\bibinfo  {journal} {Nature Reviews Physics}\ }\textbf
  {\bibinfo {volume} {5}},\ \bibinfo {pages} {157} (\bibinfo {year}
  {2023})}\BibitemShut {NoStop}%
\bibitem [{\citenamefont {Fujiwara}\ \emph {et~al.}(2020)\citenamefont
  {Fujiwara}, \citenamefont {Sun}, \citenamefont {Dohms}, \citenamefont
  {Nishimura}, \citenamefont {Suto}, \citenamefont {Takezawa}, \citenamefont
  {Oshimi}, \citenamefont {Zhao}, \citenamefont {Sadzak}, \citenamefont
  {Umehara}, \citenamefont {Teki}, \citenamefont {Komatsu}, \citenamefont
  {Benson}, \citenamefont {Shikano},\ and\ \citenamefont
  {Kage-Nakadai}}]{sensing_NV3}%
  \BibitemOpen
  \bibfield  {author} {\bibinfo {author} {\bibfnamefont {M.}~\bibnamefont
  {Fujiwara}}, \bibinfo {author} {\bibfnamefont {S.}~\bibnamefont {Sun}},
  \bibinfo {author} {\bibfnamefont {A.}~\bibnamefont {Dohms}}, \bibinfo
  {author} {\bibfnamefont {Y.}~\bibnamefont {Nishimura}}, \bibinfo {author}
  {\bibfnamefont {K.}~\bibnamefont {Suto}}, \bibinfo {author} {\bibfnamefont
  {Y.}~\bibnamefont {Takezawa}}, \bibinfo {author} {\bibfnamefont
  {K.}~\bibnamefont {Oshimi}}, \bibinfo {author} {\bibfnamefont
  {L.}~\bibnamefont {Zhao}}, \bibinfo {author} {\bibfnamefont {N.}~\bibnamefont
  {Sadzak}}, \bibinfo {author} {\bibfnamefont {Y.}~\bibnamefont {Umehara}},
  \bibinfo {author} {\bibfnamefont {Y.}~\bibnamefont {Teki}}, \bibinfo {author}
  {\bibfnamefont {N.}~\bibnamefont {Komatsu}}, \bibinfo {author} {\bibfnamefont
  {O.}~\bibnamefont {Benson}}, \bibinfo {author} {\bibfnamefont
  {Y.}~\bibnamefont {Shikano}}, \ and\ \bibinfo {author} {\bibfnamefont
  {E.}~\bibnamefont {Kage-Nakadai}},\ }\href {\doibase 10.1126/sciadv.aba9636}
  {\bibfield  {journal} {\bibinfo  {journal} {Science Advances}\ }\textbf
  {\bibinfo {volume} {6}},\ \bibinfo {pages} {eaba9636} (\bibinfo {year}
  {2020})}\BibitemShut {NoStop}%
\bibitem [{\citenamefont {Ruf}\ \emph {et~al.}(2021)\citenamefont {Ruf},
  \citenamefont {Wan}, \citenamefont {Choi}, \citenamefont {Englund},\ and\
  \citenamefont {Hanson}}]{communication_NV1}%
  \BibitemOpen
  \bibfield  {author} {\bibinfo {author} {\bibfnamefont {M.}~\bibnamefont
  {Ruf}}, \bibinfo {author} {\bibfnamefont {N.~H.}\ \bibnamefont {Wan}},
  \bibinfo {author} {\bibfnamefont {H.}~\bibnamefont {Choi}}, \bibinfo {author}
  {\bibfnamefont {D.}~\bibnamefont {Englund}}, \ and\ \bibinfo {author}
  {\bibfnamefont {R.}~\bibnamefont {Hanson}},\ }\href
  {https://pubs.aip.org/aip/jap/article/130/7/070901/1061464/Quantum-networks-based-on-color-centers-in-diamond}
  {\bibfield  {journal} {\bibinfo  {journal} {Journal of Applied Physics}\
  }\textbf {\bibinfo {volume} {130}} (\bibinfo {year} {2021})}\BibitemShut
  {NoStop}%
\bibitem [{\citenamefont {Jelezko}\ \emph
  {et~al.}(2004{\natexlab{b}})\citenamefont {Jelezko}, \citenamefont {Gaebel},
  \citenamefont {Popa}, \citenamefont {Domhan}, \citenamefont {Gruber},\ and\
  \citenamefont {Wrachtrup}}]{F.Jelezko}%
  \BibitemOpen
  \bibfield  {author} {\bibinfo {author} {\bibfnamefont {F.}~\bibnamefont
  {Jelezko}}, \bibinfo {author} {\bibfnamefont {T.}~\bibnamefont {Gaebel}},
  \bibinfo {author} {\bibfnamefont {I.}~\bibnamefont {Popa}}, \bibinfo {author}
  {\bibfnamefont {M.}~\bibnamefont {Domhan}}, \bibinfo {author} {\bibfnamefont
  {A.}~\bibnamefont {Gruber}}, \ and\ \bibinfo {author} {\bibfnamefont
  {J.}~\bibnamefont {Wrachtrup}},\ }\href {\doibase
  10.1103/PhysRevLett.93.130501} {\bibfield  {journal} {\bibinfo  {journal}
  {Physical Review Letters}\ }\textbf {\bibinfo {volume} {93}},\ \bibinfo
  {pages} {130501} (\bibinfo {year} {2004}{\natexlab{b}})}\BibitemShut
  {NoStop}%
\bibitem [{\citenamefont {Waldherr}\ \emph {et~al.}(2014)\citenamefont
  {Waldherr}, \citenamefont {Wang}, \citenamefont {Zaiser}, \citenamefont
  {Jamali}, \citenamefont {Schulte-Herbruggen}, \citenamefont {Abe},
  \citenamefont {Ohshima}, \citenamefont {Isoya}, \citenamefont {Du},
  \citenamefont {Neumann} \emph {et~al.}}]{computing_NV1}%
  \BibitemOpen
  \bibfield  {author} {\bibinfo {author} {\bibfnamefont {G.}~\bibnamefont
  {Waldherr}}, \bibinfo {author} {\bibfnamefont {Y.}~\bibnamefont {Wang}},
  \bibinfo {author} {\bibfnamefont {S.}~\bibnamefont {Zaiser}}, \bibinfo
  {author} {\bibfnamefont {M.}~\bibnamefont {Jamali}}, \bibinfo {author}
  {\bibfnamefont {T.}~\bibnamefont {Schulte-Herbruggen}}, \bibinfo {author}
  {\bibfnamefont {H.}~\bibnamefont {Abe}}, \bibinfo {author} {\bibfnamefont
  {T.}~\bibnamefont {Ohshima}}, \bibinfo {author} {\bibfnamefont
  {J.}~\bibnamefont {Isoya}}, \bibinfo {author} {\bibfnamefont
  {J.}~\bibnamefont {Du}}, \bibinfo {author} {\bibfnamefont {P.}~\bibnamefont
  {Neumann}},  \emph {et~al.},\ }\href
  {https://www.nature.com/articles/nature12919} {\bibfield  {journal} {\bibinfo
   {journal} {Nature}\ }\textbf {\bibinfo {volume} {506}},\ \bibinfo {pages}
  {204} (\bibinfo {year} {2014})}\BibitemShut {NoStop}%
\bibitem [{\citenamefont {Stas}\ \emph {et~al.}(2022)\citenamefont {Stas},
  \citenamefont {Huan}, \citenamefont {Machielse}, \citenamefont {Knall},
  \citenamefont {Suleymanzade}, \citenamefont {Pingault}, \citenamefont
  {Sutula}, \citenamefont {Ding}, \citenamefont {Knaut}, \citenamefont
  {Assumpcao} \emph {et~al.}}]{computing_NV2}%
  \BibitemOpen
  \bibfield  {author} {\bibinfo {author} {\bibfnamefont {P.-J.}\ \bibnamefont
  {Stas}}, \bibinfo {author} {\bibfnamefont {Y.~Q.}\ \bibnamefont {Huan}},
  \bibinfo {author} {\bibfnamefont {B.}~\bibnamefont {Machielse}}, \bibinfo
  {author} {\bibfnamefont {E.~N.}\ \bibnamefont {Knall}}, \bibinfo {author}
  {\bibfnamefont {A.}~\bibnamefont {Suleymanzade}}, \bibinfo {author}
  {\bibfnamefont {B.}~\bibnamefont {Pingault}}, \bibinfo {author}
  {\bibfnamefont {M.}~\bibnamefont {Sutula}}, \bibinfo {author} {\bibfnamefont
  {S.~W.}\ \bibnamefont {Ding}}, \bibinfo {author} {\bibfnamefont {C.~M.}\
  \bibnamefont {Knaut}}, \bibinfo {author} {\bibfnamefont {D.~R.}\ \bibnamefont
  {Assumpcao}},  \emph {et~al.},\ }\href
  {https://www.science.org/doi/10.1126/science.add9771} {\bibfield  {journal}
  {\bibinfo  {journal} {Science}\ }\textbf {\bibinfo {volume} {378}},\ \bibinfo
  {pages} {557} (\bibinfo {year} {2022})}\BibitemShut {NoStop}%
\bibitem [{\citenamefont {Balasubramanian}\ \emph {et~al.}(2008)\citenamefont
  {Balasubramanian}, \citenamefont {Chan}, \citenamefont {Kolesov},
  \citenamefont {Al-Hmoud}, \citenamefont {Tisler}, \citenamefont {Shin},
  \citenamefont {Kim}, \citenamefont {Wojcik}, \citenamefont {Hemmer},
  \citenamefont {Krueger}, \citenamefont {Hanke}, \citenamefont
  {Leitenstorfer}, \citenamefont {Bratschitsch}, \citenamefont {Jelezko},\ and\
  \citenamefont {Wrachtrup}}]{Balasubramanian}%
  \BibitemOpen
  \bibfield  {author} {\bibinfo {author} {\bibfnamefont {G.}~\bibnamefont
  {Balasubramanian}}, \bibinfo {author} {\bibfnamefont {I.~Y.}\ \bibnamefont
  {Chan}}, \bibinfo {author} {\bibfnamefont {R.}~\bibnamefont {Kolesov}},
  \bibinfo {author} {\bibfnamefont {M.}~\bibnamefont {Al-Hmoud}}, \bibinfo
  {author} {\bibfnamefont {J.}~\bibnamefont {Tisler}}, \bibinfo {author}
  {\bibfnamefont {C.}~\bibnamefont {Shin}}, \bibinfo {author} {\bibfnamefont
  {C.}~\bibnamefont {Kim}}, \bibinfo {author} {\bibfnamefont {A.}~\bibnamefont
  {Wojcik}}, \bibinfo {author} {\bibfnamefont {P.~R.}\ \bibnamefont {Hemmer}},
  \bibinfo {author} {\bibfnamefont {A.}~\bibnamefont {Krueger}}, \bibinfo
  {author} {\bibfnamefont {T.}~\bibnamefont {Hanke}}, \bibinfo {author}
  {\bibfnamefont {A.}~\bibnamefont {Leitenstorfer}}, \bibinfo {author}
  {\bibfnamefont {R.}~\bibnamefont {Bratschitsch}}, \bibinfo {author}
  {\bibfnamefont {F.}~\bibnamefont {Jelezko}}, \ and\ \bibinfo {author}
  {\bibfnamefont {J.}~\bibnamefont {Wrachtrup}},\ }\href {\doibase
  10.1038/nature07278} {\bibfield  {journal} {\bibinfo  {journal} {Nature}\
  }\textbf {\bibinfo {volume} {455}},\ \bibinfo {pages} {648 } (\bibinfo {year}
  {2008})}\BibitemShut {NoStop}%
\bibitem [{\citenamefont {Taylor}\ \emph {et~al.}(2008)\citenamefont {Taylor},
  \citenamefont {Cappellaro}, \citenamefont {Childress}, \citenamefont {Jiang},
  \citenamefont {Budker}, \citenamefont {Hemmer}, \citenamefont {Yacoby},
  \citenamefont {Walsworth},\ and\ \citenamefont {Lukin}}]{sensor}%
  \BibitemOpen
  \bibfield  {author} {\bibinfo {author} {\bibfnamefont {J.~M.}\ \bibnamefont
  {Taylor}}, \bibinfo {author} {\bibfnamefont {P.}~\bibnamefont {Cappellaro}},
  \bibinfo {author} {\bibfnamefont {L.}~\bibnamefont {Childress}}, \bibinfo
  {author} {\bibfnamefont {L.}~\bibnamefont {Jiang}}, \bibinfo {author}
  {\bibfnamefont {D.}~\bibnamefont {Budker}}, \bibinfo {author} {\bibfnamefont
  {P.~R.}\ \bibnamefont {Hemmer}}, \bibinfo {author} {\bibfnamefont
  {A.}~\bibnamefont {Yacoby}}, \bibinfo {author} {\bibfnamefont
  {R.}~\bibnamefont {Walsworth}}, \ and\ \bibinfo {author} {\bibfnamefont
  {M.~D.}\ \bibnamefont {Lukin}},\ }\href {\doibase 10.1038/nphys1075}
  {\bibfield  {journal} {\bibinfo  {journal} {Nature Physics}\ }\textbf
  {\bibinfo {volume} {4}},\ \bibinfo {pages} {810} (\bibinfo {year}
  {2008})}\BibitemShut {NoStop}%
\bibitem [{\citenamefont {Haque}\ and\ \citenamefont
  {Sumaiya}(2017)}]{ArifulHaque}%
  \BibitemOpen
  \bibfield  {author} {\bibinfo {author} {\bibfnamefont {A.}~\bibnamefont
  {Haque}}\ and\ \bibinfo {author} {\bibfnamefont {S.}~\bibnamefont
  {Sumaiya}},\ }\href {https://www.mdpi.com/2504-4494/1/1/6} {\bibfield
  {journal} {\bibinfo  {journal} {Journal of Manufacturing and Materials
  Processing}\ }\textbf {\bibinfo {volume} {1}} (\bibinfo {year}
  {2017})}\BibitemShut {NoStop}%
\bibitem [{\citenamefont {Gorrini}\ and\ \citenamefont
  {Bifone}(2023)}]{FedericoGorrini}%
  \BibitemOpen
  \bibfield  {author} {\bibinfo {author} {\bibfnamefont {F.}~\bibnamefont
  {Gorrini}}\ and\ \bibinfo {author} {\bibfnamefont {A.}~\bibnamefont
  {Bifone}},\ }\href {https://www.mdpi.com/2079-6374/13/7/691} {\bibfield
  {journal} {\bibinfo  {journal} {Biosensors}\ }\textbf {\bibinfo {volume}
  {13}} (\bibinfo {year} {2023})}\BibitemShut {NoStop}%
\bibitem [{\citenamefont {Luo}\ \emph {et~al.}(2022)\citenamefont {Luo},
  \citenamefont {Lindner}, \citenamefont {Langer}, \citenamefont {Cimalla},
  \citenamefont {Vidal}, \citenamefont {Hahl}, \citenamefont {Schreyvogel},
  \citenamefont {Onoda}, \citenamefont {Ishii}, \citenamefont {Ohshima},
  \citenamefont {Wang}, \citenamefont {Simpson}, \citenamefont {Johnson},
  \citenamefont {Capelli}, \citenamefont {Blinder},\ and\ \citenamefont
  {Jeske}}]{TLuo}%
  \BibitemOpen
  \bibfield  {author} {\bibinfo {author} {\bibfnamefont {T.}~\bibnamefont
  {Luo}}, \bibinfo {author} {\bibfnamefont {L.}~\bibnamefont {Lindner}},
  \bibinfo {author} {\bibfnamefont {J.}~\bibnamefont {Langer}}, \bibinfo
  {author} {\bibfnamefont {V.}~\bibnamefont {Cimalla}}, \bibinfo {author}
  {\bibfnamefont {X.}~\bibnamefont {Vidal}}, \bibinfo {author} {\bibfnamefont
  {F.}~\bibnamefont {Hahl}}, \bibinfo {author} {\bibfnamefont {C.}~\bibnamefont
  {Schreyvogel}}, \bibinfo {author} {\bibfnamefont {S.}~\bibnamefont {Onoda}},
  \bibinfo {author} {\bibfnamefont {S.}~\bibnamefont {Ishii}}, \bibinfo
  {author} {\bibfnamefont {T.}~\bibnamefont {Ohshima}}, \bibinfo {author}
  {\bibfnamefont {D.}~\bibnamefont {Wang}}, \bibinfo {author} {\bibfnamefont
  {D.~A.}\ \bibnamefont {Simpson}}, \bibinfo {author} {\bibfnamefont {B.~C.}\
  \bibnamefont {Johnson}}, \bibinfo {author} {\bibfnamefont {M.}~\bibnamefont
  {Capelli}}, \bibinfo {author} {\bibfnamefont {R.}~\bibnamefont {Blinder}}, \
  and\ \bibinfo {author} {\bibfnamefont {J.}~\bibnamefont {Jeske}},\ }\href
  {\doibase 10.1088/1367-2630/ac58b6} {\bibfield  {journal} {\bibinfo
  {journal} {New Journal of Physics}\ }\textbf {\bibinfo {volume} {24}},\
  \bibinfo {pages} {033030} (\bibinfo {year} {2022})}\BibitemShut {NoStop}%
\bibitem [{\citenamefont {Vikharev}\ \emph {et~al.}(2016)\citenamefont
  {Vikharev}, \citenamefont {Gorbachev}, \citenamefont {Lobaev}, \citenamefont
  {Muchnikov}, \citenamefont {Radishev}, \citenamefont {Isaev}, \citenamefont
  {Chernov}, \citenamefont {Bogdanov}, \citenamefont {Drozdov},\ and\
  \citenamefont {Butler}}]{deltaNVVikharev}%
  \BibitemOpen
  \bibfield  {author} {\bibinfo {author} {\bibfnamefont {A.~L.}\ \bibnamefont
  {Vikharev}}, \bibinfo {author} {\bibfnamefont {A.~M.}\ \bibnamefont
  {Gorbachev}}, \bibinfo {author} {\bibfnamefont {M.~A.}\ \bibnamefont
  {Lobaev}}, \bibinfo {author} {\bibfnamefont {A.~B.}\ \bibnamefont
  {Muchnikov}}, \bibinfo {author} {\bibfnamefont {D.~B.}\ \bibnamefont
  {Radishev}}, \bibinfo {author} {\bibfnamefont {V.~A.}\ \bibnamefont {Isaev}},
  \bibinfo {author} {\bibfnamefont {V.~V.}\ \bibnamefont {Chernov}}, \bibinfo
  {author} {\bibfnamefont {S.~A.}\ \bibnamefont {Bogdanov}}, \bibinfo {author}
  {\bibfnamefont {M.~N.}\ \bibnamefont {Drozdov}}, \ and\ \bibinfo {author}
  {\bibfnamefont {J.~E.}\ \bibnamefont {Butler}},\ }\href {\doibase
  10.1002/pssr.201510453} {\bibfield  {journal} {\bibinfo  {journal} {Physica
  Status Solidi RRL}\ }\textbf {\bibinfo {volume} {10}},\ \bibinfo {pages}
  {324–327} (\bibinfo {year} {2016})}\BibitemShut {NoStop}%
\bibitem [{\citenamefont {Bogdanov}\ \emph {et~al.}(2020)\citenamefont
  {Bogdanov}, \citenamefont {Bolshedvorskii}, \citenamefont {Zeleneev},
  \citenamefont {Soshenko}, \citenamefont {Rubinas}, \citenamefont {Radishev},
  \citenamefont {Lobaev}, \citenamefont {Vikharev}, \citenamefont {Gorbachev},
  \citenamefont {Drozdov}, \citenamefont {Sorokin},\ and\ \citenamefont
  {Akimov}}]{deltaNVBogdanov}%
  \BibitemOpen
  \bibfield  {author} {\bibinfo {author} {\bibfnamefont {S.}~\bibnamefont
  {Bogdanov}}, \bibinfo {author} {\bibfnamefont {S.}~\bibnamefont
  {Bolshedvorskii}}, \bibinfo {author} {\bibfnamefont {A.}~\bibnamefont
  {Zeleneev}}, \bibinfo {author} {\bibfnamefont {V.}~\bibnamefont {Soshenko}},
  \bibinfo {author} {\bibfnamefont {O.}~\bibnamefont {Rubinas}}, \bibinfo
  {author} {\bibfnamefont {D.}~\bibnamefont {Radishev}}, \bibinfo {author}
  {\bibfnamefont {M.}~\bibnamefont {Lobaev}}, \bibinfo {author} {\bibfnamefont
  {A.}~\bibnamefont {Vikharev}}, \bibinfo {author} {\bibfnamefont
  {A.}~\bibnamefont {Gorbachev}}, \bibinfo {author} {\bibfnamefont
  {M.}~\bibnamefont {Drozdov}}, \bibinfo {author} {\bibfnamefont
  {V.}~\bibnamefont {Sorokin}}, \ and\ \bibinfo {author} {\bibfnamefont
  {A.}~\bibnamefont {Akimov}},\ }\href {\doibase 10.1016/j.mtcomm.2020.101019}
  {\bibfield  {journal} {\bibinfo  {journal} {Materials Today Communications}\
  }\textbf {\bibinfo {volume} {24}},\ \bibinfo {pages} {101019} (\bibinfo
  {year} {2020})}\BibitemShut {NoStop}%
\bibitem [{\citenamefont {Pezzagna}\ \emph
  {et~al.}(2010{\natexlab{a}})\citenamefont {Pezzagna}, \citenamefont
  {Naydenov}, \citenamefont {Jelezko}, \citenamefont {Wrachtrup},\ and\
  \citenamefont {Meijer}}]{Jelezko3}%
  \BibitemOpen
  \bibfield  {author} {\bibinfo {author} {\bibfnamefont {S.}~\bibnamefont
  {Pezzagna}}, \bibinfo {author} {\bibfnamefont {B.}~\bibnamefont {Naydenov}},
  \bibinfo {author} {\bibfnamefont {F.}~\bibnamefont {Jelezko}}, \bibinfo
  {author} {\bibfnamefont {J.}~\bibnamefont {Wrachtrup}}, \ and\ \bibinfo
  {author} {\bibfnamefont {J.}~\bibnamefont {Meijer}},\ }\href {\doibase
  10.1088/1367-2630/12/6/065017} {\bibfield  {journal} {\bibinfo  {journal}
  {New Journal of Physics}\ }\textbf {\bibinfo {volume} {12}},\ \bibinfo
  {pages} {065017} (\bibinfo {year} {2010}{\natexlab{a}})}\BibitemShut
  {NoStop}%
\bibitem [{\citenamefont {Acosta}\ and\ \citenamefont
  {Hemmer}(2013)}]{AcostaHemmer}%
  \BibitemOpen
  \bibfield  {author} {\bibinfo {author} {\bibfnamefont {V.}~\bibnamefont
  {Acosta}}\ and\ \bibinfo {author} {\bibfnamefont {P.}~\bibnamefont
  {Hemmer}},\ }\href {\doibase 10.1557/mrs.2013.18} {\bibfield  {journal}
  {\bibinfo  {journal} {MRS Bulletin}\ }\textbf {\bibinfo {volume} {38}},\
  \bibinfo {pages} {127–130} (\bibinfo {year} {2013})}\BibitemShut {NoStop}%
\bibitem [{\citenamefont {Addhya}\ \emph {et~al.}(2024)\citenamefont {Addhya},
  \citenamefont {Tyne}, \citenamefont {Guo}, \citenamefont {Hammock},
  \citenamefont {Li}, \citenamefont {Leung}, \citenamefont {DeVault},
  \citenamefont {Awschalom}, \citenamefont {Delegan}, \citenamefont
  {Heremans},\ and\ \citenamefont {High}}]{laser_writting}%
  \BibitemOpen
  \bibfield  {author} {\bibinfo {author} {\bibfnamefont {A.}~\bibnamefont
  {Addhya}}, \bibinfo {author} {\bibfnamefont {V.}~\bibnamefont {Tyne}},
  \bibinfo {author} {\bibfnamefont {X.}~\bibnamefont {Guo}}, \bibinfo {author}
  {\bibfnamefont {I.~N.}\ \bibnamefont {Hammock}}, \bibinfo {author}
  {\bibfnamefont {Z.}~\bibnamefont {Li}}, \bibinfo {author} {\bibfnamefont
  {M.}~\bibnamefont {Leung}}, \bibinfo {author} {\bibfnamefont {C.~T.}\
  \bibnamefont {DeVault}}, \bibinfo {author} {\bibfnamefont {D.~D.}\
  \bibnamefont {Awschalom}}, \bibinfo {author} {\bibfnamefont {N.}~\bibnamefont
  {Delegan}}, \bibinfo {author} {\bibfnamefont {F.~J.}\ \bibnamefont
  {Heremans}}, \ and\ \bibinfo {author} {\bibfnamefont {A.~A.}\ \bibnamefont
  {High}},\ }\href {\doibase 10.1021/acs.nanolett.4c02639} {\bibfield
  {journal} {\bibinfo  {journal} {Nano Letters}\ }\textbf {\bibinfo {volume}
  {24}},\ \bibinfo {pages} {11224} (\bibinfo {year} {2024})}\BibitemShut
  {NoStop}%
\bibitem [{\citenamefont {Hausmann}\ \emph {et~al.}(2010)\citenamefont
  {Hausmann}, \citenamefont {Khan}, \citenamefont {Zhang}, \citenamefont
  {Babinec}, \citenamefont {Martinick}, \citenamefont {McCutcheon},
  \citenamefont {Hemmer},\ and\ \citenamefont {Lon\v{c}ar}}]{nanowire}%
  \BibitemOpen
  \bibfield  {author} {\bibinfo {author} {\bibfnamefont {B.~J.}\ \bibnamefont
  {Hausmann}}, \bibinfo {author} {\bibfnamefont {M.}~\bibnamefont {Khan}},
  \bibinfo {author} {\bibfnamefont {Y.}~\bibnamefont {Zhang}}, \bibinfo
  {author} {\bibfnamefont {T.~M.}\ \bibnamefont {Babinec}}, \bibinfo {author}
  {\bibfnamefont {K.}~\bibnamefont {Martinick}}, \bibinfo {author}
  {\bibfnamefont {M.}~\bibnamefont {McCutcheon}}, \bibinfo {author}
  {\bibfnamefont {P.~R.}\ \bibnamefont {Hemmer}}, \ and\ \bibinfo {author}
  {\bibfnamefont {M.}~\bibnamefont {Lon\v{c}ar}},\ }\href {\doibase
  https://doi.org/10.1016/j.diamond.2010.01.011} {\bibfield  {journal}
  {\bibinfo  {journal} {Diamond and Related Materials}\ }\textbf {\bibinfo
  {volume} {19}},\ \bibinfo {pages} {621} (\bibinfo {year} {2010})}\BibitemShut
  {NoStop}%
\bibitem [{\citenamefont {Irber}\ \emph {et~al.}(2021)\citenamefont {Irber},
  \citenamefont {Kong}, \citenamefont {Kieschnick}, \citenamefont {Lühmann},
  \citenamefont {Kwiatkowski}, \citenamefont {Meijer}, \citenamefont {Du},
  \citenamefont {Shi},\ and\ \citenamefont {Reinhard}}]{DominikIrber}%
  \BibitemOpen
  \bibfield  {author} {\bibinfo {author} {\bibfnamefont {F.}~\bibnamefont
  {Irber}, \bibfnamefont {Dominik M.and~Poggiali}}, \bibinfo {author}
  {\bibfnamefont {F.}~\bibnamefont {Kong}}, \bibinfo {author} {\bibfnamefont
  {M.}~\bibnamefont {Kieschnick}}, \bibinfo {author} {\bibfnamefont
  {T.}~\bibnamefont {Lühmann}}, \bibinfo {author} {\bibfnamefont
  {D.}~\bibnamefont {Kwiatkowski}}, \bibinfo {author} {\bibfnamefont
  {J.}~\bibnamefont {Meijer}}, \bibinfo {author} {\bibfnamefont
  {J.}~\bibnamefont {Du}}, \bibinfo {author} {\bibfnamefont {F.}~\bibnamefont
  {Shi}}, \ and\ \bibinfo {author} {\bibfnamefont {F.}~\bibnamefont
  {Reinhard}},\ }\href {https://www.nature.com/articles/s41467-020-20755-3}
  {\bibfield  {journal} {\bibinfo  {journal} {Nature Communications}\ }\textbf
  {\bibinfo {volume} {12}} (\bibinfo {year} {2021})}\BibitemShut {NoStop}%
\bibitem [{\citenamefont {fei Meng}\ \emph {et~al.}(2008)\citenamefont {fei
  Meng}, \citenamefont {shiue Yan}, \citenamefont {Lai}, \citenamefont
  {Krasnicki}, \citenamefont {Shu}, \citenamefont {Yu}, \citenamefont {Liang},
  \citenamefont {kwang Mao},\ and\ \citenamefont {Hemley}}]{annealingLPHT}%
  \BibitemOpen
  \bibfield  {author} {\bibinfo {author} {\bibfnamefont {Y.}~\bibnamefont {fei
  Meng}}, \bibinfo {author} {\bibfnamefont {C.}~\bibnamefont {shiue Yan}},
  \bibinfo {author} {\bibfnamefont {J.}~\bibnamefont {Lai}}, \bibinfo {author}
  {\bibfnamefont {S.}~\bibnamefont {Krasnicki}}, \bibinfo {author}
  {\bibfnamefont {H.}~\bibnamefont {Shu}}, \bibinfo {author} {\bibfnamefont
  {T.}~\bibnamefont {Yu}}, \bibinfo {author} {\bibfnamefont {Q.}~\bibnamefont
  {Liang}}, \bibinfo {author} {\bibfnamefont {H.}~\bibnamefont {kwang Mao}}, \
  and\ \bibinfo {author} {\bibfnamefont {R.~J.}\ \bibnamefont {Hemley}},\
  }\href {\doibase 10.1073/pnas.0808230105} {\bibfield  {journal} {\bibinfo
  {journal} {Proceedings of the National Academy of Sciences}\ }\textbf
  {\bibinfo {volume} {105}},\ \bibinfo {pages} {17620} (\bibinfo {year}
  {2008})}\BibitemShut {NoStop}%
\bibitem [{\citenamefont {Naydenov}\ \emph {et~al.}(2010)\citenamefont
  {Naydenov}, \citenamefont {Reinhard}, \citenamefont {Lämmle}, \citenamefont
  {Richter}, \citenamefont {Kalish}, \citenamefont {D’Haenens-Johansson},
  \citenamefont {Newton}, \citenamefont {Jelezko},\ and\ \citenamefont
  {Wrachtrup}}]{annealing1}%
  \BibitemOpen
  \bibfield  {author} {\bibinfo {author} {\bibfnamefont {B.}~\bibnamefont
  {Naydenov}}, \bibinfo {author} {\bibfnamefont {F.}~\bibnamefont {Reinhard}},
  \bibinfo {author} {\bibfnamefont {A.}~\bibnamefont {Lämmle}}, \bibinfo
  {author} {\bibfnamefont {V.}~\bibnamefont {Richter}}, \bibinfo {author}
  {\bibfnamefont {R.}~\bibnamefont {Kalish}}, \bibinfo {author} {\bibfnamefont
  {U.~F.~S.}\ \bibnamefont {D’Haenens-Johansson}}, \bibinfo {author}
  {\bibfnamefont {M.}~\bibnamefont {Newton}}, \bibinfo {author} {\bibfnamefont
  {F.}~\bibnamefont {Jelezko}}, \ and\ \bibinfo {author} {\bibfnamefont
  {J.}~\bibnamefont {Wrachtrup}},\ }\href {\doibase 10.1063/1.3527975}
  {\bibfield  {journal} {\bibinfo  {journal} {Applied Physics Letters}\
  }\textbf {\bibinfo {volume} {97}},\ \bibinfo {pages} {242511} (\bibinfo
  {year} {2010})}\BibitemShut {NoStop}%
\bibitem [{\citenamefont {Pezzagna}\ \emph
  {et~al.}(2010{\natexlab{b}})\citenamefont {Pezzagna}, \citenamefont
  {Naydenov}, \citenamefont {Jelezko}, \citenamefont {Wrachtrup},\ and\
  \citenamefont {Meijer}}]{annealing2}%
  \BibitemOpen
  \bibfield  {author} {\bibinfo {author} {\bibfnamefont {S.}~\bibnamefont
  {Pezzagna}}, \bibinfo {author} {\bibfnamefont {B.}~\bibnamefont {Naydenov}},
  \bibinfo {author} {\bibfnamefont {F.}~\bibnamefont {Jelezko}}, \bibinfo
  {author} {\bibfnamefont {J.}~\bibnamefont {Wrachtrup}}, \ and\ \bibinfo
  {author} {\bibfnamefont {J.}~\bibnamefont {Meijer}},\ }\href {\doibase
  10.1088/1367-2630/12/6/065017} {\bibfield  {journal} {\bibinfo  {journal}
  {New Journal of Physics}\ }\textbf {\bibinfo {volume} {12}},\ \bibinfo
  {pages} {065017} (\bibinfo {year} {2010}{\natexlab{b}})}\BibitemShut
  {NoStop}%
\bibitem [{\citenamefont {L{\"u}hmann}\ \emph {et~al.}(2019)\citenamefont
  {L{\"u}hmann}, \citenamefont {John}, \citenamefont {Wunderlich},
  \citenamefont {Meijer},\ and\ \citenamefont {Pezzagna}}]{charge_diffusion}%
  \BibitemOpen
  \bibfield  {author} {\bibinfo {author} {\bibfnamefont {T.}~\bibnamefont
  {L{\"u}hmann}}, \bibinfo {author} {\bibfnamefont {R.}~\bibnamefont {John}},
  \bibinfo {author} {\bibfnamefont {R.}~\bibnamefont {Wunderlich}}, \bibinfo
  {author} {\bibfnamefont {J.}~\bibnamefont {Meijer}}, \ and\ \bibinfo {author}
  {\bibfnamefont {S.}~\bibnamefont {Pezzagna}},\ }\href
  {https://www.nature.com/articles/s41467-019-12556-0} {\bibfield  {journal}
  {\bibinfo  {journal} {Nature communications}\ }\textbf {\bibinfo {volume}
  {10}},\ \bibinfo {pages} {4956} (\bibinfo {year} {2019})}\BibitemShut
  {NoStop}%
\bibitem [{\citenamefont {Jaffe}\ \emph {et~al.}(2020)\citenamefont {Jaffe},
  \citenamefont {Attrash}, \citenamefont {Kuntumalla}, \citenamefont
  {Akhvlediani}, \citenamefont {Michaelson}, \citenamefont {Gal}, \citenamefont
  {Felgen}, \citenamefont {Fischer}, \citenamefont {Reithmaier}, \citenamefont
  {Popov}, \citenamefont {Hoffman},\ and\ \citenamefont
  {Orenstein}}]{technion}%
  \BibitemOpen
  \bibfield  {author} {\bibinfo {author} {\bibfnamefont {T.}~\bibnamefont
  {Jaffe}}, \bibinfo {author} {\bibfnamefont {M.}~\bibnamefont {Attrash}},
  \bibinfo {author} {\bibfnamefont {M.~K.}\ \bibnamefont {Kuntumalla}},
  \bibinfo {author} {\bibfnamefont {R.}~\bibnamefont {Akhvlediani}}, \bibinfo
  {author} {\bibfnamefont {S.}~\bibnamefont {Michaelson}}, \bibinfo {author}
  {\bibfnamefont {L.}~\bibnamefont {Gal}}, \bibinfo {author} {\bibfnamefont
  {N.}~\bibnamefont {Felgen}}, \bibinfo {author} {\bibfnamefont
  {M.}~\bibnamefont {Fischer}}, \bibinfo {author} {\bibfnamefont {J.~P.}\
  \bibnamefont {Reithmaier}}, \bibinfo {author} {\bibfnamefont
  {C.}~\bibnamefont {Popov}}, \bibinfo {author} {\bibfnamefont
  {A.}~\bibnamefont {Hoffman}}, \ and\ \bibinfo {author} {\bibfnamefont
  {M.}~\bibnamefont {Orenstein}},\ }\href {\doibase
  10.1021/acs.nanolett.9b05243} {\bibfield  {journal} {\bibinfo  {journal}
  {Nano Letters}\ }\textbf {\bibinfo {volume} {20}},\ \bibinfo {pages} {3192}
  (\bibinfo {year} {2020})}\BibitemShut {NoStop}%
\bibitem [{\citenamefont {Kuntumalla}\ \emph {et~al.}(2021)\citenamefont
  {Kuntumalla}, \citenamefont {Attrash}, \citenamefont {Fischer}, \citenamefont
  {Michaelson}, \citenamefont {Kravchuk},\ and\ \citenamefont
  {Hoffman}}]{technion2}%
  \BibitemOpen
  \bibfield  {author} {\bibinfo {author} {\bibfnamefont {M.~K.}\ \bibnamefont
  {Kuntumalla}}, \bibinfo {author} {\bibfnamefont {M.}~\bibnamefont {Attrash}},
  \bibinfo {author} {\bibfnamefont {M.}~\bibnamefont {Fischer}}, \bibinfo
  {author} {\bibfnamefont {S.}~\bibnamefont {Michaelson}}, \bibinfo {author}
  {\bibfnamefont {T.}~\bibnamefont {Kravchuk}}, \ and\ \bibinfo {author}
  {\bibfnamefont {A.}~\bibnamefont {Hoffman}},\ }\href {\doibase
  https://doi.org/10.1016/j.apsusc.2021.149331} {\bibfield  {journal} {\bibinfo
   {journal} {Applied Surface Science}\ }\textbf {\bibinfo {volume} {550}},\
  \bibinfo {pages} {149331} (\bibinfo {year} {2021})}\BibitemShut {NoStop}%
\bibitem [{\citenamefont {Cui}\ \emph {et~al.}(2015)\citenamefont {Cui},
  \citenamefont {Greenspon}, \citenamefont {Ohno}, \citenamefont {Myers},
  \citenamefont {Jayich}, \citenamefont {Awschalom},\ and\ \citenamefont
  {Hu}}]{ICPdamageNVnearSurface}%
  \BibitemOpen
  \bibfield  {author} {\bibinfo {author} {\bibfnamefont {S.}~\bibnamefont
  {Cui}}, \bibinfo {author} {\bibfnamefont {A.~S.}\ \bibnamefont {Greenspon}},
  \bibinfo {author} {\bibfnamefont {K.}~\bibnamefont {Ohno}}, \bibinfo {author}
  {\bibfnamefont {B.~A.}\ \bibnamefont {Myers}}, \bibinfo {author}
  {\bibfnamefont {A.~C.~B.}\ \bibnamefont {Jayich}}, \bibinfo {author}
  {\bibfnamefont {D.~D.}\ \bibnamefont {Awschalom}}, \ and\ \bibinfo {author}
  {\bibfnamefont {E.~L.}\ \bibnamefont {Hu}},\ }\href {\doibase
  10.1021/acs.nanolett.5b00457} {\bibfield  {journal} {\bibinfo  {journal}
  {Nano Letters}\ }\textbf {\bibinfo {volume} {15}},\ \bibinfo {pages} {2887}
  (\bibinfo {year} {2015})}\BibitemShut {NoStop}%
\bibitem [{\citenamefont {Kim}\ \emph {et~al.}(2014)\citenamefont {Kim},
  \citenamefont {Mamin}, \citenamefont {Sherwood}, \citenamefont {Rettner},
  \citenamefont {Frommer},\ and\ \citenamefont
  {Rugar}}]{ICPdamageNVnearSurface2}%
  \BibitemOpen
  \bibfield  {author} {\bibinfo {author} {\bibfnamefont {M.}~\bibnamefont
  {Kim}}, \bibinfo {author} {\bibfnamefont {H.~J.}\ \bibnamefont {Mamin}},
  \bibinfo {author} {\bibfnamefont {M.~H.}\ \bibnamefont {Sherwood}}, \bibinfo
  {author} {\bibfnamefont {C.~T.}\ \bibnamefont {Rettner}}, \bibinfo {author}
  {\bibfnamefont {J.}~\bibnamefont {Frommer}}, \ and\ \bibinfo {author}
  {\bibfnamefont {D.}~\bibnamefont {Rugar}},\ }\href {\doibase
  10.1063/1.4891839} {\bibfield  {journal} {\bibinfo  {journal} {Applied
  Physics Letters}\ }\textbf {\bibinfo {volume} {105}},\ \bibinfo {pages}
  {042406} (\bibinfo {year} {2014})}\BibitemShut {NoStop}%
\bibitem [{\citenamefont {F\'avaro~de Oliveira}\ \emph
  {et~al.}(2015)\citenamefont {F\'avaro~de Oliveira}, \citenamefont
  {Momenzadeh}, \citenamefont {Wang}, \citenamefont {Konuma}, \citenamefont
  {Markham}, \citenamefont {Edmonds}, \citenamefont {Denisenko},\ and\
  \citenamefont {Wrachtrup}}]{ICPdamageNVnearSurface3}%
  \BibitemOpen
  \bibfield  {author} {\bibinfo {author} {\bibfnamefont {F.}~\bibnamefont
  {F\'avaro~de Oliveira}}, \bibinfo {author} {\bibfnamefont {S.~A.}\
  \bibnamefont {Momenzadeh}}, \bibinfo {author} {\bibfnamefont
  {Y.}~\bibnamefont {Wang}}, \bibinfo {author} {\bibfnamefont {M.}~\bibnamefont
  {Konuma}}, \bibinfo {author} {\bibfnamefont {M.}~\bibnamefont {Markham}},
  \bibinfo {author} {\bibfnamefont {A.~M.}\ \bibnamefont {Edmonds}}, \bibinfo
  {author} {\bibfnamefont {A.}~\bibnamefont {Denisenko}}, \ and\ \bibinfo
  {author} {\bibfnamefont {J.}~\bibnamefont {Wrachtrup}},\ }\href {\doibase
  10.1063/1.4929356} {\bibfield  {journal} {\bibinfo  {journal} {Applied
  Physics Letters}\ }\textbf {\bibinfo {volume} {107}},\ \bibinfo {pages}
  {073107} (\bibinfo {year} {2015})}\BibitemShut {NoStop}%
\bibitem [{\citenamefont {Tsunaki}\ \emph
  {et~al.}(2024{\natexlab{a}})\citenamefont {Tsunaki}, \citenamefont {Singh},
  \citenamefont {Volkova}, \citenamefont {Trofimov}, \citenamefont
  {Pregnolato}, \citenamefont {Schröder},\ and\ \citenamefont
  {Naydenov}}]{setup_1}%
  \BibitemOpen
  \bibfield  {author} {\bibinfo {author} {\bibfnamefont {L.}~\bibnamefont
  {Tsunaki}}, \bibinfo {author} {\bibfnamefont {A.}~\bibnamefont {Singh}},
  \bibinfo {author} {\bibfnamefont {K.}~\bibnamefont {Volkova}}, \bibinfo
  {author} {\bibfnamefont {S.}~\bibnamefont {Trofimov}}, \bibinfo {author}
  {\bibfnamefont {T.}~\bibnamefont {Pregnolato}}, \bibinfo {author}
  {\bibfnamefont {T.}~\bibnamefont {Schröder}}, \ and\ \bibinfo {author}
  {\bibfnamefont {B.}~\bibnamefont {Naydenov}},\ }\href
  {https://arxiv.org/abs/2407.09411} {\bibfield  {journal} {\bibinfo  {journal}
  {arXiv}\ }\textbf {\bibinfo {volume} {quant-ph}} (\bibinfo {year}
  {2024}{\natexlab{a}})}\BibitemShut {NoStop}%
\bibitem [{\citenamefont {Panadero}\ \emph {et~al.}(2024)\citenamefont
  {Panadero}, \citenamefont {Espin\'os}, \citenamefont {Tsunaki}, \citenamefont
  {Volkova}, \citenamefont {Tobalina}, \citenamefont {Casanova}, \citenamefont
  {Acedo}, \citenamefont {Naydenov}, \citenamefont {Puebla},\ and\
  \citenamefont {Torrontegui}}]{setup_2}%
  \BibitemOpen
  \bibfield  {author} {\bibinfo {author} {\bibfnamefont {I.}~\bibnamefont
  {Panadero}}, \bibinfo {author} {\bibfnamefont {H.}~\bibnamefont {Espin\'os}},
  \bibinfo {author} {\bibfnamefont {L.}~\bibnamefont {Tsunaki}}, \bibinfo
  {author} {\bibfnamefont {K.}~\bibnamefont {Volkova}}, \bibinfo {author}
  {\bibfnamefont {A.}~\bibnamefont {Tobalina}}, \bibinfo {author}
  {\bibfnamefont {J.}~\bibnamefont {Casanova}}, \bibinfo {author}
  {\bibfnamefont {P.}~\bibnamefont {Acedo}}, \bibinfo {author} {\bibfnamefont
  {B.}~\bibnamefont {Naydenov}}, \bibinfo {author} {\bibfnamefont
  {R.}~\bibnamefont {Puebla}}, \ and\ \bibinfo {author} {\bibfnamefont
  {E.}~\bibnamefont {Torrontegui}},\ }\href {\doibase
  10.1103/PhysRevApplied.22.014035} {\bibfield  {journal} {\bibinfo  {journal}
  {Physical Review Applied}\ }\textbf {\bibinfo {volume} {22}},\ \bibinfo
  {pages} {014035} (\bibinfo {year} {2024})}\BibitemShut {NoStop}%
\bibitem [{\citenamefont {Volkova}\ \emph {et~al.}(2022)\citenamefont
  {Volkova}, \citenamefont {Heupel}, \citenamefont {Trofimov}, \citenamefont
  {Betz}, \citenamefont {Colom}, \citenamefont {MacQueen}, \citenamefont
  {Akhundzada}, \citenamefont {Reginka}, \citenamefont {Ehresmann},
  \citenamefont {Reithmaier}, \citenamefont {Burger}, \citenamefont {Popov},\
  and\ \citenamefont {Naydenov}}]{nanopillars_kseniia}%
  \BibitemOpen
  \bibfield  {author} {\bibinfo {author} {\bibfnamefont {K.}~\bibnamefont
  {Volkova}}, \bibinfo {author} {\bibfnamefont {J.}~\bibnamefont {Heupel}},
  \bibinfo {author} {\bibfnamefont {S.}~\bibnamefont {Trofimov}}, \bibinfo
  {author} {\bibfnamefont {F.}~\bibnamefont {Betz}}, \bibinfo {author}
  {\bibfnamefont {R.}~\bibnamefont {Colom}}, \bibinfo {author} {\bibfnamefont
  {R.~W.}\ \bibnamefont {MacQueen}}, \bibinfo {author} {\bibfnamefont
  {S.}~\bibnamefont {Akhundzada}}, \bibinfo {author} {\bibfnamefont
  {M.}~\bibnamefont {Reginka}}, \bibinfo {author} {\bibfnamefont
  {A.}~\bibnamefont {Ehresmann}}, \bibinfo {author} {\bibfnamefont {J.~P.}\
  \bibnamefont {Reithmaier}}, \bibinfo {author} {\bibfnamefont
  {S.}~\bibnamefont {Burger}}, \bibinfo {author} {\bibfnamefont
  {C.}~\bibnamefont {Popov}}, \ and\ \bibinfo {author} {\bibfnamefont
  {B.}~\bibnamefont {Naydenov}},\ }\href
  {https://www.mdpi.com/2079-4991/12/9/1516} {\bibfield  {journal} {\bibinfo
  {journal} {Nanomaterials}\ }\textbf {\bibinfo {volume} {12}} (\bibinfo {year}
  {2022})}\BibitemShut {NoStop}%
\bibitem [{\citenamefont {Binder}\ \emph {et~al.}(2017)\citenamefont {Binder},
  \citenamefont {Stark}, \citenamefont {Tomek}, \citenamefont {Scheuer},
  \citenamefont {Frank}, \citenamefont {Jahnke}, \citenamefont {Müller},
  \citenamefont {Schmitt}, \citenamefont {Metsch}, \citenamefont {Unden},
  \citenamefont {Gehring}, \citenamefont {Huck}, \citenamefont {Andersen},
  \citenamefont {Rogers},\ and\ \citenamefont {Jelezko}}]{qudi}%
  \BibitemOpen
  \bibfield  {author} {\bibinfo {author} {\bibfnamefont {J.~M.}\ \bibnamefont
  {Binder}}, \bibinfo {author} {\bibfnamefont {A.}~\bibnamefont {Stark}},
  \bibinfo {author} {\bibfnamefont {N.}~\bibnamefont {Tomek}}, \bibinfo
  {author} {\bibfnamefont {J.}~\bibnamefont {Scheuer}}, \bibinfo {author}
  {\bibfnamefont {F.}~\bibnamefont {Frank}}, \bibinfo {author} {\bibfnamefont
  {K.~D.}\ \bibnamefont {Jahnke}}, \bibinfo {author} {\bibfnamefont
  {C.}~\bibnamefont {Müller}}, \bibinfo {author} {\bibfnamefont
  {S.}~\bibnamefont {Schmitt}}, \bibinfo {author} {\bibfnamefont {M.~H.}\
  \bibnamefont {Metsch}}, \bibinfo {author} {\bibfnamefont {T.}~\bibnamefont
  {Unden}}, \bibinfo {author} {\bibfnamefont {T.}~\bibnamefont {Gehring}},
  \bibinfo {author} {\bibfnamefont {A.}~\bibnamefont {Huck}}, \bibinfo {author}
  {\bibfnamefont {U.~L.}\ \bibnamefont {Andersen}}, \bibinfo {author}
  {\bibfnamefont {L.~J.}\ \bibnamefont {Rogers}}, \ and\ \bibinfo {author}
  {\bibfnamefont {F.}~\bibnamefont {Jelezko}},\ }\href {\doibase
  https://doi.org/10.1016/j.softx.2017.02.001} {\bibfield  {journal} {\bibinfo
  {journal} {SoftwareX}\ }\textbf {\bibinfo {volume} {6}},\ \bibinfo {pages}
  {85} (\bibinfo {year} {2017})}\BibitemShut {NoStop}%
\bibitem [{\citenamefont {Petkov}\ \emph {et~al.}(2013)\citenamefont {Petkov},
  \citenamefont {Rendler}, \citenamefont {Petkov}, \citenamefont {Schnabel},
  \citenamefont {Reithmaier}, \citenamefont {Wrachtrup}, \citenamefont
  {Popov},\ and\ \citenamefont {Kulisch}}]{raman_diamond}%
  \BibitemOpen
  \bibfield  {author} {\bibinfo {author} {\bibfnamefont {E.}~\bibnamefont
  {Petkov}}, \bibinfo {author} {\bibfnamefont {T.}~\bibnamefont {Rendler}},
  \bibinfo {author} {\bibfnamefont {C.}~\bibnamefont {Petkov}}, \bibinfo
  {author} {\bibfnamefont {F.}~\bibnamefont {Schnabel}}, \bibinfo {author}
  {\bibfnamefont {J.~P.}\ \bibnamefont {Reithmaier}}, \bibinfo {author}
  {\bibfnamefont {J.}~\bibnamefont {Wrachtrup}}, \bibinfo {author}
  {\bibfnamefont {C.}~\bibnamefont {Popov}}, \ and\ \bibinfo {author}
  {\bibfnamefont {W.}~\bibnamefont {Kulisch}},\ }\href {\doibase
  https://doi.org/10.1002/pssa.201329282} {\bibfield  {journal} {\bibinfo
  {journal} {physica status solidi (a)}\ }\textbf {\bibinfo {volume} {210}},\
  \bibinfo {pages} {2066} (\bibinfo {year} {2013})}\BibitemShut {NoStop}%
\bibitem [{\citenamefont {Siyushev}\ \emph {et~al.}(2010)\citenamefont
  {Siyushev}, \citenamefont {Kaiser}, \citenamefont {Jacques}, \citenamefont
  {Gerhardt}, \citenamefont {Bischof}, \citenamefont {Fedder}, \citenamefont
  {Dodson}, \citenamefont {Markham}, \citenamefont {Twitchen}, \citenamefont
  {Jelezko} \emph {et~al.}}]{sat_function}%
  \BibitemOpen
  \bibfield  {author} {\bibinfo {author} {\bibfnamefont {P.}~\bibnamefont
  {Siyushev}}, \bibinfo {author} {\bibfnamefont {F.}~\bibnamefont {Kaiser}},
  \bibinfo {author} {\bibfnamefont {V.}~\bibnamefont {Jacques}}, \bibinfo
  {author} {\bibfnamefont {I.}~\bibnamefont {Gerhardt}}, \bibinfo {author}
  {\bibfnamefont {S.}~\bibnamefont {Bischof}}, \bibinfo {author} {\bibfnamefont
  {H.}~\bibnamefont {Fedder}}, \bibinfo {author} {\bibfnamefont
  {J.}~\bibnamefont {Dodson}}, \bibinfo {author} {\bibfnamefont
  {M.}~\bibnamefont {Markham}}, \bibinfo {author} {\bibfnamefont
  {D.}~\bibnamefont {Twitchen}}, \bibinfo {author} {\bibfnamefont
  {F.}~\bibnamefont {Jelezko}},  \emph {et~al.},\ }\href
  {https://pubs.aip.org/aip/apl/article/97/24/241902/122677/Monolithic-diamond-optics-for-single-photon}
  {\bibfield  {journal} {\bibinfo  {journal} {Applied physics letters}\
  }\textbf {\bibinfo {volume} {97}} (\bibinfo {year} {2010})}\BibitemShut
  {NoStop}%
\bibitem [{\citenamefont {Hahn}(1950)}]{hahn}%
  \BibitemOpen
  \bibfield  {author} {\bibinfo {author} {\bibfnamefont {E.~L.}\ \bibnamefont
  {Hahn}},\ }\href {\doibase 10.1103/PhysRev.80.580} {\bibfield  {journal}
  {\bibinfo  {journal} {Physical Review}\ }\textbf {\bibinfo {volume} {80}},\
  \bibinfo {pages} {580} (\bibinfo {year} {1950})}\BibitemShut {NoStop}%
\bibitem [{\citenamefont {Rabi}\ \emph {et~al.}(1939)\citenamefont {Rabi},
  \citenamefont {Millman}, \citenamefont {Kusch},\ and\ \citenamefont
  {Zacharias}}]{rabi}%
  \BibitemOpen
  \bibfield  {author} {\bibinfo {author} {\bibfnamefont {I.~I.}\ \bibnamefont
  {Rabi}}, \bibinfo {author} {\bibfnamefont {S.}~\bibnamefont {Millman}},
  \bibinfo {author} {\bibfnamefont {P.}~\bibnamefont {Kusch}}, \ and\ \bibinfo
  {author} {\bibfnamefont {J.~R.}\ \bibnamefont {Zacharias}},\ }\href {\doibase
  10.1103/PhysRev.55.526} {\bibfield  {journal} {\bibinfo  {journal} {Physical
  Review}\ }\textbf {\bibinfo {volume} {55}},\ \bibinfo {pages} {526} (\bibinfo
  {year} {1939})}\BibitemShut {NoStop}%
\bibitem [{\citenamefont {Tsunaki}\ \emph
  {et~al.}(2024{\natexlab{b}})\citenamefont {Tsunaki}, \citenamefont
  {Bauerhenne}, \citenamefont {Xibraku}, \citenamefont {Garcia}, \citenamefont
  {Singer},\ and\ \citenamefont {Naydenov}}]{qtoken_1}%
  \BibitemOpen
  \bibfield  {author} {\bibinfo {author} {\bibfnamefont {L.}~\bibnamefont
  {Tsunaki}}, \bibinfo {author} {\bibfnamefont {B.}~\bibnamefont {Bauerhenne}},
  \bibinfo {author} {\bibfnamefont {M.}~\bibnamefont {Xibraku}}, \bibinfo
  {author} {\bibfnamefont {M.~E.}\ \bibnamefont {Garcia}}, \bibinfo {author}
  {\bibfnamefont {K.}~\bibnamefont {Singer}}, \ and\ \bibinfo {author}
  {\bibfnamefont {B.}~\bibnamefont {Naydenov}},\ }\href
  {https://arxiv.org/abs/2412.08530} {\bibfield  {journal} {\bibinfo  {journal}
  {arXiv}\ }\textbf {\bibinfo {volume} {quant-ph}} (\bibinfo {year}
  {2024}{\natexlab{b}})}\BibitemShut {NoStop}%
\bibitem [{\citenamefont {Bauerhenne}\ \emph {et~al.}(2024)\citenamefont
  {Bauerhenne}, \citenamefont {Tsunaki}, \citenamefont {Thieme}, \citenamefont
  {Naydenov},\ and\ \citenamefont {Singer}}]{qtoken_2}%
  \BibitemOpen
  \bibfield  {author} {\bibinfo {author} {\bibfnamefont {B.}~\bibnamefont
  {Bauerhenne}}, \bibinfo {author} {\bibfnamefont {L.}~\bibnamefont {Tsunaki}},
  \bibinfo {author} {\bibfnamefont {J.}~\bibnamefont {Thieme}}, \bibinfo
  {author} {\bibfnamefont {B.}~\bibnamefont {Naydenov}}, \ and\ \bibinfo
  {author} {\bibfnamefont {K.}~\bibnamefont {Singer}},\ }\href
  {https://arxiv.org/abs/2412.20243} {\bibfield  {journal} {\bibinfo  {journal}
  {arXiv}\ }\textbf {\bibinfo {volume} {quant-ph}} (\bibinfo {year}
  {2024})}\BibitemShut {NoStop}%
\bibitem [{\citenamefont {Li}\ \emph {et~al.}(2021)\citenamefont {Li},
  \citenamefont {Zheng}, \citenamefont {Peng}, \citenamefont {Kamiya},
  \citenamefont {Niki}, \citenamefont {Stepanov}, \citenamefont {Jarmola},
  \citenamefont {Shimizu}, \citenamefont {Takahashi}, \citenamefont
  {Wickenbrock},\ and\ \citenamefont {Budker}}]{DEER}%
  \BibitemOpen
  \bibfield  {author} {\bibinfo {author} {\bibfnamefont {S.}~\bibnamefont
  {Li}}, \bibinfo {author} {\bibfnamefont {H.}~\bibnamefont {Zheng}}, \bibinfo
  {author} {\bibfnamefont {Z.}~\bibnamefont {Peng}}, \bibinfo {author}
  {\bibfnamefont {M.}~\bibnamefont {Kamiya}}, \bibinfo {author} {\bibfnamefont
  {T.}~\bibnamefont {Niki}}, \bibinfo {author} {\bibfnamefont {V.}~\bibnamefont
  {Stepanov}}, \bibinfo {author} {\bibfnamefont {A.}~\bibnamefont {Jarmola}},
  \bibinfo {author} {\bibfnamefont {Y.}~\bibnamefont {Shimizu}}, \bibinfo
  {author} {\bibfnamefont {S.}~\bibnamefont {Takahashi}}, \bibinfo {author}
  {\bibfnamefont {A.}~\bibnamefont {Wickenbrock}}, \ and\ \bibinfo {author}
  {\bibfnamefont {D.}~\bibnamefont {Budker}},\ }\href {\doibase
  10.1103/PhysRevB.104.094307} {\bibfield  {journal} {\bibinfo  {journal}
  {Physical Review B}\ }\textbf {\bibinfo {volume} {104}},\ \bibinfo {pages}
  {094307} (\bibinfo {year} {2021})}\BibitemShut {NoStop}%
\end{thebibliography}%

\end{document}